\documentclass[11pt,journal,onecolumn,romanappendices]{IEEEtran}

\usepackage[utf8]{inputenc} 
\usepackage[T1]{fontenc}
\usepackage{url}
\usepackage{ifthen}
\usepackage{cite}
\usepackage[cmex10]{amsmath} 
\usepackage{longtable}
\usepackage{makecell}
\usepackage{algorithm,algorithmic}
\usepackage{graphicx}
\usepackage{rotating}
\usepackage[bottom]{footmisc}
                             
\usepackage{amssymb}
\usepackage{amsthm}
\usepackage{booktabs}
\usepackage{xifthen,xparse}
\usepackage{hyperref}  
\usepackage{float}
\usepackage{bbm}
\usepackage[percent]{overpic}

\usepackage{tikz,pgfplots,tkz-graph,color}
\pgfplotsset{compat=newest}
\pgfplotsset{plot coordinates/math parser=false}
\usetikzlibrary{decorations.markings,calc,positioning,matrix,spy,shapes.misc}

\usepackage{textcomp}

\allowdisplaybreaks

\newcommand\independent{\protect\mathpalette{\protect\independenT}{\perp}}
\def\independenT#1#2{\mathrel{\rlap{$#1#2$}\mkern2mu{#1#2}}}
\newtheorem{theorem}{Theorem}
\newtheorem{corollary}[theorem]{Corollary}

\newtheorem{lemma}[theorem]{Lemma}
\newtheorem{definition}[theorem]{Definition}

\newtheorem{example}{Example}

\newenvironment{example*}
  {\addtocounter{example}{-1}\example}
  {\endexample}

\newcommand{\ket}[1]{\left\lvert #1 \right\rangle}
\newcommand{\bra}[1]{\left\langle #1 \right\rvert}
\NewDocumentCommand\ketbra{+m+g}{%
  \IfNoValueTF{#2}
    {\left\lvert #1 \right\rangle \left\langle #1 \right\vert}
  {\left\lvert #1 \right\rangle \left\langle #2 \right\rvert}%
}
\NewDocumentCommand\braket{+m+g}{%
  \IfNoValueTF{#2}
    {\left\langle #1 \vert #1 \right\rangle}
  {\left\langle #1 \vert #2 \right\rangle}%
}

\newcommand{\rhohat}{\rho}

\newcommand{\trays}[1]{\text{Tr}\left[ #1 \right]}

\newcommand{\abs}[1]{|#1|}
\newcommand{\rdbrs}[1]{(#1)}
\newcommand{\clbrs}[1]{\{#1 \}}
\newcommand{\clbrsv}[2]{\{#1| #2 \}}
\newcommand{\rdbrsv}[2]{(#1 | #2)}
\newcommand{\Tr}[1]{\text{Tr}(#1)}

\usepackage{xcolor}
\definecolor{purp}{RGB}{160, 32, 240}
\definecolor{lightblue}{RGB}{66, 134, 244}

\title{Adaptive Procedures for Discrimination Between \\ Arbitrary Tensor-Product Quantum States}

 \author{%
   \IEEEauthorblockN{Sarah Brandsen\IEEEauthorrefmark{1}$^1$,
                     Mengke Lian\IEEEauthorrefmark{2}$^1$,
                     Kevin D. Stubbs\IEEEauthorrefmark{3},
                     Narayanan Rengaswamy\IEEEauthorrefmark{2},
                     and Henry D. Pfister\IEEEauthorrefmark{2}}%
   \thanks{$^1$These authors contributed equally to this research.
           S. Brandsen is with the 
           Department of Physics, 
           Duke University, 
           Durham, North Carolina 27708, USA.
           M. Lian, N. Rengaswamy, and H. D. Pfister are with the
           Department of Electrical and Computer Engineering,
           Duke University,
           Durham, North Carolina 27708, USA.
           K. D. Stubbs is with the
           Department of Mathematics,
           Duke University, 
           Durham, North Carolina 27708, USA.
           Correspondence Email: sarah.brandsen@duke.edu%
  }}

\begin{document}

\maketitle

\begin{abstract}
Discrimination between quantum states is a fundamental task in quantum information theory. Given two arbitrary tensor-product quantum states (TPQS) $\rhohat_{\pm} = \rhohat_{\pm}^{(1)} \otimes \cdots \otimes \rhohat_{\pm}^{(N)}$, determining the joint $N$-system measurement to optimally distinguish between the two states is a hard problem ~\cite{Holevo-book}. Thus, there is great interest in identifying local measurement schemes that are optimal or close-to-optimal. In this work, we focus on distinguishing between two general TPQS. We begin by generalizing previous work by Acin et al. (Phys. Rev. A \textbf{71}, 032338) to show that a locally greedy (LG) scheme using Bayesian updating can optimally distinguish between two states that can be written as tensor products of arbitrary pure states. Then, we show that even in the limit of large $N$ the same algorithm cannot distinguish tensor products of mixed states with vanishing error probability. This poor asymptotic behavior occurs because the Helstrom measurement becomes trivial for sufficiently biased priors. Based on this, we introduce a modified locally greedy (MLG) scheme with strictly better performance. 

In the second part of this work, we compare these simple local schemes with a general dynamic programming (DP) approach that finds the optimal series of local measurements to distinguish the two states. When the subsystems are non-identical, we demonstrate that the ordering of the systems affects performance and we extend the DP technique to determine the optimal ordering adaptively. Finally, in contrast to the binary optimal collective measurement, we show that adaptive protocols on sufficiently large (e.g., qutrit) subsystems must contain non-binary measurements to be optimal. (The code that produced the simulation results in this paper can be found at: \url{https://github.com/SarahBrandsen/AdaptiveStateDiscrimination})
\end{abstract}

\section{Introduction}
\label{sec:intro}

Measurement lies at the heart of quantum mechanics. 
Since the exact state of a quantum system cannot be directly observed, measurement is the primary means of understanding real quantum systems ~\cite{Peres, Renes, Bisio, Dall'Arno, Busch, Spekkens}. However, due to the inherent uncertainty in quantum systems it is impossible to design a quantum measurement capable of perfectly discriminating between two non-orthogonal quantum states ~\cite{Nielsen-2010, Gisin}. 
The optimal measurement for state discrimination was described by Helstrom~\cite{Helstrom-jsp69}. 
However, for composite quantum systems, the Helstrom measurement may be computationally expensive to solve and impractical to implement experimentally as it requires simultaneously measuring all subsystems.

Several works in the literature have investigated techniques that use only local operations to distinguish between two possible qubit states, given $N$ copies of the state, with the aim of achieving or approximating the Helstrom probability of success.  
The simplest strategy, a na{\"i}ve ``majority vote'', has been shown to have probability of error which approaches zero exponentially fast in $N$~\cite{Acin-physreva05,Higgins-physreva11}. 
Furthermore, for the special case when all copies are pure states, it has been shown that a greedy adaptive strategy involving Bayesian updates of the prior after each measurement result is optimal~\cite{Acin-physreva05}. 
Finally, dynamic programming has been utilized to recursively minimize the expected future error over all possible allowed measurements, and thus yields the optimal adaptive strategy for any given family of measurements~\cite{Higgins-physreva11}. 

There has also been much work in the direction of unambiguous discrimination, where we allow three outcomes: ``first state with certainty'', ``second state with certainty'', or ``not sure'' ~\cite{Bergou-jpcs07, Eldar2, Eldar3}. Several works also have considered the task of distinguishing between $m>2$ possible quantum states, although for general multi-state discrimination problems there is no known optimal solution ~\cite{Eldar, Yuen, Helstrom2, Wootters94}.  

In this paper, we generalize previous works and consider the problem of discrimination between two arbitrary TPQS with a focus on qubit and qutrit subsystems. 
More specifically, we suppose that we are given either $\rhohat_+$ or $\rhohat_-$ with prior probability $q$ and $1-q$ respectively, where $\rhohat_\pm = \rho^{(1)}_\pm \otimes \cdots \otimes \rho^{(N)}_\pm$ and $\rho^{(j)}_\pm$ is potentially different for each $j \in \{ 1, \cdots, N \}$.  This problem is of practical interest in quantum communications, where we might modulate a classical binary codeword into a TPQS in order to transmit information through multiple uses of the channel, and each subsystem could experience a (slightly) different channel parameter. 
(From a more theoretical perspective, proving results about optimal adaptive discrimination between states of this form admit some technical machinery to perform inductive arguments.)

When distinct systems are in different states, the optimal measurement order for the subsystems depends on the measurement outcome of the previous subsystems.
We prove that, if all of the systems are pure states, then the order of measurement does not matter and a locally greedy Bayesian update-based strategy is optimal.
This generalizes the result in~\cite{Acin-physreva05} mentioned above. 

When the states are mixed, the locally greedy algorithm is no longer optimal and in fact performs worse than most nonadaptive local strategies in the limit as $N \rightarrow \infty$. We show that this poor asymptotic performance arises from the local Helstrom measurement becoming noninformative for sufficiently imbalanced priors. To overcome this, we introduce a modified locally greedy adaptive strategy with strictly better performance. 

We also discuss a dynamic programming-based strategy that finds the optimal locally adaptive strategy. (A closely related technique is described in~\cite{Higgins-physreva11}. This dynamic programming approach is the optimal locally adaptive technique subject to some simple constraints and includes the locally greedy techniques as a special case of itself.)

Finally, we consider the performance of ternary and binary projective measurements over qutrit states and show that, in general, multiple-outcome measurements are needed for optimality. This holds even for a binary state discrimination problem. 
Numerical results are provided for all these scenarios and the source code used to generate them is available at \url{https://github.com/SarahBrandsen/AdaptiveStateDiscrimination}.

\section{Main Contributions}
\label{sec:main_contributions}

Many works in the literature consider only the case when the $N$ subsystems are copies of the same state. For example, this occurs when one is allowed to use the same state multiple times in order to perform the discrimination.
The primary difference in our work is that we consider tensor products of subsystems with potentially distinct states and dimension and provide insights into the factors that affect performance in this setting, beyond knowledge of the two possibilities alone.

We will consider locally adaptive schemes that measure local subsystems in each round and determine parameters of the next measurement as a function of the past measurement results.
We will discuss the following factors and their impact on the overall performance of such schemes:
\begin{enumerate}

\item The order in which the subsystems are measured (unless $\rhohat_{\pm}^{(i)} = \rhohat_{\pm}^{(j)}\ \forall\ i, j \in \{1,\ldots,N\}$ so that the subsystems are identical copies).

\item The algorithm which provides the adaptive measurement for the $j$-th subsystem given all previous measurement results. 
The algorithms we consider are (in order of complexity):
\begin{enumerate}

\item \emph{Locally Greedy and Modified Locally Greedy} (referred to as ``locally optimal locally adaptive''~\cite{Higgins-physreva11}): 
After each measurement, one updates the prior probability using Bayes' theorem. 
If we denote all measurement results before the $j$-th subsystem as $\mathbf{d}_{1:j-1}$, or more succinctly $\mathbf{d}_{[j-1]}$ with $[j-1] \triangleq \{ 1, \ldots, j-1 \}$, and the updated prior given these measurements as $P_{j}(q, \mathbf{d}_{[j-1]})$, the measurement implemented on the $j$-th subsystem is then the Helstrom measurement for $\{ (P_{j}(q, \mathbf{d}_{[j-1]}), \rhohat_{+}^{(j)}), (1- P_{j}(q, \mathbf{d}_{[j-1]}), \rhohat_{-}^{(j)})\}$. We additionally consider a ``modified locally greedy'' (MLG) variation of this method which differs from the locally greedy method only when the local Helstrom measurement for the updated prior is trivial (namely, equal to the zero or identity operator.)

\item \emph{Dynamic Programming} (referred to as ``globally optimal locally adaptive'' in ~\cite{Higgins-physreva11}): 
Let $\sigma \in \mathcal{S}_N$, where $\mathcal{S}_N$ denotes the symmetric group on $N$ elements, represent a permutation that provides the index of the subsystem to be measured at every round.
Initially $\sigma$ is unknown, and the algorithm progressively defines the permutation index by index.
At the $j$-th round, the scheme chooses both the subsystem $\sigma(j)$ to be measured next and the actual measurement as follows.
Given the results for the first $j-1$ measurements (on subsystems $\sigma(1), \ldots, \sigma(j-1)$), the subsystem $\sigma(j)$ and the measurement on it are chosen using a recursive expression for expected future risk, i.e., probability of error from future actions over all possible orderings of the remaining subsystems, so as to minimize the expected error over all possible outputs. 
Dynamic programming is the optimal adaptive technique, given a family of allowable measurements, including all other local techniques as special cases of itself.
We refer to this method as the \emph{Measurement- and Order-Optimized DYnamic} (MOODY) algorithm, where the measurements are optimized over the family of orthogonal projective measurements.

\end{enumerate}


\item The number of measurement outcomes allowed (e.g. binary measurements versus ternary measurements).

\end{enumerate}

We summarize our results below:

\begin{enumerate}

\item[(i)] For the case where $\rhohat_{\pm}$ is a tensor product of arbitrary pure states, we analytically prove that the locally greedy algorithm (and hence the MOODY algorithm) achieves the optimal Helstrom probability of success.

\begin{theorem}
\label{thm:pure_states}
Let $P_{\mathrm{s, h}}(q, \rhohat_{\pm})$ and $P_{\mathrm{s, lg}}(q, \rhohat_{\pm})$ denote the probabilities of successfully discriminating the states $\rhohat_{+}$ and $\rhohat_{-}$ using the joint $N$-system Helstrom measurement and the locally greedy measurement technique respectively, given initial prior $\mathbb{P}(\rhohat = \rhohat_{+}) = q$ . 
If $\rhohat_{+}$ and $\rhohat_{-}$ are pure states, i.e., $\rhohat_{\pm}^{(j)} \triangleq \ketbra{\theta_{\pm}^{(j)}}$ for some $\theta_{\pm}^{(j)} \in (0, 2\pi)$ for every $j \in [N]$, where $\ket{\theta_{\pm}} \triangleq \cos \frac{\theta}{2} \ket{0} \pm \sin \frac{\theta}{2} \ket{1}$, then
\begin{align}
P_{\mathrm{s, h}}(q, \rhohat_{\pm}) &= P_{\mathrm{s, lg}}(q, \rhohat_{\pm}) = \frac{1}{2} \left( 1 + \sqrt{1 - \Pi_{j=0}^{N} \cos^{2}(\theta_{j})} \, \right).
\end{align}
Here we have defined the overlaps between states in the zeroth subsystem to be 
$\theta_{0} \triangleq \sin^{-1}(2q-1)$.
\end{theorem}

{\em Sketch of Proof:} The strategy is to prove the result for $N=2$ and then extend via induction for arbitrary $N$. 
\begin{flushright}
See Appendix~\ref{sec:pure_states_proof} for the complete proof.            \ \ \ \ \ \ \ \ \ \ \ \ \ \ \ \ \ \ \ \ \ \ \ \ \ \ \ \ \ \ \  \ \ \ \ \ \ \ \ \ \ \ \ \ \ \ \ \ \ \ \ \ \ \ \ \  \ \ \ \ \ \ \ \ \ \ \ \ \ \ \ \ \ \ \ \ \ \ \  
$\blacksquare$
\end{flushright}

\item[(ii)] When $\rhohat_{\pm}$ are tensor products of depolarized pure states, with depolarizing parameter $\gamma$, we have the following empirical results:
\begin{enumerate}

\item{ This occurs both when the subsystems are identical copies and when the subsystems are distinct. First, we plot the probability of success of the locally greedy and modified locally greedy algorithms and observe a plateau in the performance for increasing $N$ under the locally greedy algorithm.
We prove an upper bound for the success probability as a function of the prior $q$ and the channel depolarizing parameter, namely, $P_{\text{succ}, \gamma}(\rhohat_{\pm}, q) \leq \max \big \{ q, 1-q, \frac{(1 - \frac{\gamma}{2})^{2}}{(1 - \frac{\gamma}{2})^{2} + (\frac{\gamma}{2})^{2}} \big\}$.}
\item{ We also introduce an improved version of the locally greedy algorithm. This improved algorithm performs at least as well as the locally greedy algorithm for distinguishing between arbitrary QTPS $\rho_{+}$ and $\rho_{-}$. Furthermore, its success probability converges to unity for large $N$.}

\item We plot the difference in success probability between the best ordering of subsystems and the worst ordering of subsystems. The ordering is determined by the order-optimized locally greedy and MOODY algorithms.

\end{enumerate} 


\item[(iii)] We plot the probability of success of the MOODY algorithm as a function of the depolarizing parameter $\gamma$ for depolarized pure qutrit states (using orthogonal projective binary and ternary measurements).

\begin{enumerate}

\item From the difference in success probabilities between the best and worst ordering, we observe that there is still a non-trivial effect of ordering beyond qubit states (for several values of $N$).

\item From the difference in probabilities of success between ternary adaptive measurements and binary adaptive measurements, we observe that ternary measurements perform better.
Thus, even for a binary state discrimination problem, we see that the size of the measurement set for optimal performance can be larger than $2$.

\end{enumerate}

\end{enumerate}

The remainder of this paper is structured as follows.
The notation is defined in Section~\ref{sec:notation}, and the locally greedy and modified locally greedy algorithms are described in Section~\ref{sec:locally_greedy}.
Subsequently, we discuss dynamic programming-based adaptive approaches in Section~\ref{sec:dp_algorithms} and provide empirical results mentioned above.
Finally, we conclude the paper in Section~\ref{sec:conclusion} by summarizing our contributions and discussing planned future work.

\section{Notation}
\label{sec:notation}

Following the same notation as above, $\rhohat$ is the random variable representing the given state, so that either $\rhohat = \rhohat_{+} = \rhohat_{+}^{(1)} \otimes \cdots \otimes \rhohat_{+}^{(N)}$ or $\rhohat = \rhohat_{-} = \rhohat_{-}^{(1)} \otimes \cdots \otimes \rhohat_{-}^{(N)}$, and we refer to $N$ as the number of subsystems. 
In general, each $\rhohat_{\pm}^{(j)}$ is an arbitrary density matrix of finite dimensions, but we consider only qubits and qutrits in this paper. We additionally require that each $\rhohat_{\pm}^{(j)}$ be a real matrix. In the qubit case, we use the parametrization $\rho_{\pm}^{(j)} \triangleq (1-\gamma) \ket{\theta_{\pm, j}} \bra{\theta_{\pm, j}} + \frac{\gamma}{2} \mathbb{I}$, where $\ket{\theta} \triangleq \cos{\frac{\theta}{2}} \ket{0} + \sin{\frac{\theta}{2}} \ket{1}$ and $j \in \{1, 2, .., N\}$ denotes the subsystem index.


The prior probability of state $\rhohat_{+}$ is denoted by $q \triangleq \mathbb{P}[\rhohat = \rhohat_{+}]$.
The number of subsystems measured jointly in each round is denoted by $m \in \{ 1,\ldots,N \}$, where $m$ divides $N$, and $m = 1$ unless otherwise mentioned.
The permutation $\sigma \in \mathcal{S}_N$, where $\mathcal{S}_N$ is the symmetric group on $N$ elements, is unknown at the beginning of the protocol, and is defined progressively in each round (index by index) when the algorithm determines the next subsystem to measure (assuming no grouping of subsystems, i.e., $m = 1$).
At round $j \in \{1,\ldots,N\}$, we determine the next subsystem $\sigma(j)$. We denote by $A_{\sigma(j)}$ the random variable corresponding to the action which takes values $\mathbf{a}_{\sigma(j)} \in \mathcal{A}$ (once given all previous measurement results, $A_{\sigma(j)}$ is deterministic and is found by optimizing a cost function). The measurement result upon executing the action can be represented by the random variable $D_{\sigma(j)}$ which takes values $d_{\sigma(j)} \in \mathcal{D}$. Here $\mathcal{A}$ is a generic action set which is specified by the type of measurements in any specific scheme, and $\mathcal{D}$ is the space containing possible outcomes for the chosen action set.
For example, if $\mathcal{A}$ contains projective measurements on qubits, then $\mathcal{D} = \{ \pm 1 \}$.
For a natural number $n$, define $[n] \triangleq \{ 1,\ldots, n \}$.
Then at round $j$, the past actions and results are recorded into the vectors $\mathbf{a}_{[j-1]}^{\sigma} = (\mathbf{a}_{\sigma(1)}, \ldots, \mathbf{a}_{\sigma(j-1)})$ and $\mathbf{d}_{[j-1]}^{\sigma} = (d_{\sigma(1)}, \ldots, d_{\sigma(j-1)})$ respectively.

\section{Locally Greedy Algorithm}
\label{sec:locally_greedy}

First, we describe a simple locally greedy algorithm, which was called the ``locally optimal locally adaptive'' algorithm in~\cite{Higgins-physreva11}.
For $m = 1$, at each round $j \in [N]$, the algorithm updates the probability that the given state is $\rhohat_{+}$ based on results of past measurements.
The algorithm does not consider any non-trivial ordering of the $N$ subsystems, so $\sigma(j) = j$ for all $j \in [N]$.
Once the prior is updated at round $j$, it performs the Helstrom measurement on the subsystem $j$ according to the given $\rhohat_{\pm}^{(j)}$ and this updated prior.
In order to formally describe this process and later generalize it to the dynamic-programming algorithm in the next section, we begin by defining the conditional state probability at round $j$ for a non-trivial permutation $\sigma$ on the $N$ subsystems.

\begin{definition}
The \emph{conditional state probability (CSP)} $C_{j}^{\sigma}(q, \mathbf{a}_{[j]}^{\sigma}, \mathbf{d}_{[j]}^{\sigma})$ is defined as the probability that $\rhohat = \rhohat_{+}$ given that the starting prior was $q$, that the first $j$ rounds of measurement were executed with ordering $\sigma$ and actions $\mathbf{a}_{[j]}^{\sigma}$, and that the results were $\mathbf{d}_{[j]}^{\sigma}$. Therefore, the updated prior at round $j$ is the corresponding CSP 
\begin{align}
C_{j}^{\sigma}(q, \mathbf{a}_{[j]}^{\sigma}, \mathbf{d}_{[j]}^{\sigma}) \triangleq \mathbb{P}\left( \rhohat = \rhohat_{+} \biggr\vert\  A_{[j]}^{\sigma} = \mathbf{a}_{[j]}^{\sigma},\ D_{[j]}^{\sigma} = \mathbf{d}_{[j]}^{\sigma} \right).
\end{align}
Thus, when $j=0$ we recover the initial prior as $C_{0}^{\sigma}(q) \triangleq q$. The dependence of the conditional probability on the initial prior $q$ is left implicit in the above definition and in the following. 
\end{definition}

Then, the CSP can be computed using past actions and results as
\begin{align}
  &  C_j^\sigma (q, \mathbf{a}_{[j]}^\sigma, \mathbf{d}_{[j]}^\sigma )  \\    
    & = \frac{ \mathbb{P}\left( \rhohat_{+}, d_{\sigma(j)} \biggr\vert \mathbf{a}_{[j]}^\sigma, \mathbf{d}_{[j-1]}^\sigma \right) }{ \mathbb{P}\left( d_{\sigma(j)} \biggr\vert \mathbf{a}_{[j]}^\sigma, \mathbf{d}_{[j-1]}^\sigma \right)}  \\
    & = \frac{ \mathbb{P}\left( \rhohat_{+}, d_{\sigma(j)} \biggr\vert \mathbf{a}_{[j]}^\sigma, \mathbf{d}_{[j-1]}^\sigma \right)}{ \mathbb{P}\left( \rhohat_{+}, d_{\sigma(j)} \biggr\vert \mathbf{a}_{[j]}^\sigma, \mathbf{d}_{[j-1]}^\sigma \right) + \mathbb{P}\left( \rhohat_{-}, d_{\sigma(j)} \biggr\vert \mathbf{a}_{[j]}^\sigma, \mathbf{d}_{[j-1]}^\sigma \right) },  \\
   & = \frac{ \mathbb{P}\left( d_{\sigma(j)} \biggr\vert \rhohat_{+}, \mathbf{a}_{[j]}^\sigma, \mathbf{d}_{[j-1]}^\sigma \right) \mathbb{P}\left( \rhohat_{+} \biggr\vert \mathbf{a}_{[j]}^\sigma, \mathbf{d}_{[j-1]}^\sigma \right) }{\mathbb{P}\left( d_{\sigma(j)} \biggr\vert \rhohat_{+}, \mathbf{a}_{[j]}^\sigma, \mathbf{d}_{[j-1]}^\sigma \right)
 \mathbb{P}\left( \rhohat_{+} \biggr\vert \mathbf{a}_{[j]}^\sigma, \mathbf{d}_{[j-1]}^\sigma \right)
 + \mathbb{P}\left( d_{\sigma(j)} \biggr\vert \rhohat_{-}, \mathbf{a}_{[j]}^\sigma, \mathbf{d}_{[j-1]}^\sigma \right) \mathbb{P}\left( \rhohat_{-} \biggr\vert \mathbf{a}_{[j]}^\sigma, \mathbf{d}_{[j-1]}^\sigma \right)},
\end{align}
where, in the second equality, we have marginalized over the possible values of $\rhohat$ in the denominator.
Now we make some observations that will allow us to simplify this expression.
\begin{itemize}

\item Since all measurements are performed on different subsystems, the outcome of the $j$-th measurement does not depend on the previous $j - 1$ measurements once the $j$-th action is given, i.e.,
\begin{align}
( d_{\sigma(j)} \independent (q, \mathbf{a}_{[j-1]}^\sigma, \mathbf{d}_{[j-1]}^\sigma) ) \,\mid\, \mathbf{a}_{\sigma(j)} \Longrightarrow \mathbb{P} ( d_{\sigma(j)} | \rhohat_\pm, \mathbf{a}_{[j]}^\sigma, \mathbf{d}_{[j-1]}^\sigma ) = \mathbb{P} ( d_{\sigma(j)} | \rhohat_\pm, \mathbf{a}_{\sigma(j)})
\end{align}

\item Executing a measurement action without knowing the outcome does not help with inference, namely
\begin{align}
\mathbb{P}\left( \rhohat = \rhohat_{+} \biggr\vert \mathbf{a}_{[j]}^\sigma, \mathbf{d}_{[j-1]}^\sigma \right)
  & = \mathbb{P}\left( \rhohat = \rhohat_{+} \biggr\vert \mathbf{a}_{[j-1]}^\sigma, \mathbf{d}_{[j-1]}^\sigma \right)
    = C_{j-1}^\sigma ( q, \mathbf{a}_{[j-1]}^\sigma, \mathbf{d}_{[j-1]}^\sigma ), \\
\mathbb{P} \left( \rhohat = \rhohat_{-} \biggr\vert \mathbf{a}_{[j]}^\sigma, \mathbf{d}_{[j-1]}^\sigma \right)
  & = \mathbb{P} \left( \rhohat = \rhohat_{-} \biggr\vert \mathbf{a}_{[j-1]}^\sigma, \mathbf{d}_{[j-1]}^\sigma \right)
    = 1 - C_{j-1}^\sigma ( q, \mathbf{a}_{[j-1]}^\sigma, \mathbf{d}_{[j-1]}^\sigma ).
\end{align}
        
\item For two different permutations $\sigma$ and $\tau$, if for all $i = 1, \ldots, j$ we have
$\sigma(i) = \tau(i), \ \mathbf{a}_{\sigma(i)} = \mathbf{a}_{\tau(i)}, \ \text{and}\ d_{\sigma(i)} = d_{\tau(i)}$, 
then
$C_j^\sigma ( q, \mathbf{a}_{[j]}^\sigma, \mathbf{d}_{[j]}^\sigma ) = C_j^\tau ( q, \mathbf{a}_{[j]}^\tau, \mathbf{d}_{[j]}^\tau )$.
        
\end{itemize}
Applying these observations in the expression for $C_{j}^{\sigma}(q, \mathbf{a}_{[j]}^{\sigma}, \mathbf{d}_{[j]}^{\sigma})$, we have
\begin{align}
C_j^\sigma ( q, \mathbf{a}_{[j]}^\sigma, d_{[j]}^\sigma )        
    & = \frac{ \mathbb{P} \left( d_{\sigma(j)} \biggr\vert \rhohat_{+}, \mathbf{a}_{\sigma(j)} \right) C_{j-1} ( q, \mathbf{a}_{[j-1]}^\sigma, \mathbf{d}_{[j-1]}^\sigma ) }{ \mathbb{P} \left( d_{\sigma(j)} \biggr\vert \rhohat_{+}, \mathbf{a}_{\sigma(j)} \right) C_{j-1}^\sigma ( q, \mathbf{a}_{[j-1]}^\sigma, \mathbf{d}_{[j-1]}^\sigma ) + \mathbb{P} \left( d_{\sigma(j)} \biggr\vert \rhohat_{-}, \mathbf{a}_{\sigma(j)} \right) \left( 1 - C_{j-1}^\sigma ( q, \mathbf{a}_{[j-1]}^\sigma, \mathbf{d}_{[j-1]}^\sigma ) \right) }.  
\end{align}
Now we observe that, in the recursion, only the term $C_{j-1}^\sigma (q, \mathbf{a}_{[j-1]}^\sigma, \mathbf{d}_{[j-1]}^\sigma)$ involves the variables $q, \mathbf{a}_{[j-1]}^\sigma, \mathbf{d}_{[j-1]}^\sigma$. 
Thus, we can simplify the notation by defining two quantities:
\begin{align}
\mathcal{L}(p, \mathbf{a}, d) \triangleq \mathbb{P}\left( d \mid \rhohat_{+}, \mathbf{a} \right) \cdot p + \mathbb{P}\left( d \mid \rhohat_{-}, \mathbf{a} \right) \cdot (1 - p), \qquad
\mathcal{P}(p, \mathbf{a}, d) \triangleq \frac{\mathbb{P} \left( d \mid \rhohat_{+}, \mathbf{a} \right) \cdot p}{\mathcal{L}(p, \mathbf{a}, d)}.
\end{align}
The naming follows from observing that they represent a likelihood and a posterior, respectively.
Thus we can write 
\begin{align}
\mathbb{P} \left( d_{\sigma(j)} \biggr\vert \mathbf{a}_{[j]}^\sigma, \mathbf{d}_{[j-1]}^\sigma \right) 
  & = \mathcal{L} (C_{j-1}^\sigma ( q, \mathbf{a}_{[j-1]}^\sigma, \mathbf{d}_{[j-1]}^\sigma ), \mathbf{a}_{\sigma(j)}, d_{\sigma(j)}) ,
    \label{top_down_recursion_denominator_order} \\
C_j^\sigma ( q, \mathbf{a}_{[j]}^\sigma, \mathbf{d}_{[j]}^\sigma ) 
  & = \mathcal{P} (C_{j-1}^\sigma ( q, \mathbf{a}_{[j-1]}^\sigma, \mathbf{d}_{[j-1]}^\sigma ), \mathbf{a}_{\sigma(j)}, d_{\sigma(j)}).
    \label{top_down_recursion_order}
\end{align}
This completes the description of the Bayesian update in the locally greedy algorithm.

Next we discuss the performance of this algorithm when the $N$ subsystems are identical copies of qubits. In this case, the ordering of the subsystems is clearly immaterial.
At round $j$, the locally greedy algorithm uses the CSP $C_{j-1}^{\sigma}(q, \mathbf{a}_{[j-1]}^{\sigma}, \mathbf{d}_{[j-1]}^{\sigma})$ and applies the (optimal) Helstrom measurement on the $j$-th subsystem.
This measurement is defined by the projector
\begin{align}
\label{eq:helstrom_j}
\Pi(p, j) \triangleq \sum_{\ket{v} \in \mathcal{V}(p, j)} \ketbra{v}{v}, \ \text{where}\ \mathcal{V}(p, j) \triangleq \{ \ket{v} \colon ((1-p) \rhohat_{-}^{(j)} - p \rhohat_{+}^{(j)})  \ket{v} = \lambda \ket{v}\ \text{and}\ \lambda \geq 0 \},
\end{align}
where $p = C_{j-1}^{\sigma}(q, \mathbf{a}_{[j-1]}^{\sigma}, \mathbf{d}_{[j-1]}^{\sigma})$.
Since $\rhohat_{\pm}^{(i)} = \rhohat_{\pm}^{(j)}$ for all $i,j \in [N]$, $\Pi(p,j)$ changes at every round only because of the changing prior $p$.
The outcome probabilities for this measurement are given by
\begin{align}
\label{eq:outcome_prob_j}
\mathbb{P} \left( d \mid \rhohat_\pm^{(j)}, \Pi(p, j) \right) = 
\begin{cases}
1 - \Tr{\Pi(p, j) \rhohat_\pm^{(j)}} & \text{ if } d = +1, \\
\Tr{\Pi(p, j) \rhohat_\pm^{(j)}}     & \text{ if } d = -1,
\end{cases}
\end{align}
and the overall probability of error (at round $j$) is given by
\begin{align}
\label{eq:prob_error_j}
P_{\mathrm{err}, j} = (1 - \Tr{\Pi(p, j) \rhohat_{-}^{(j)}}) \cdot (1-p) + \Tr{\Pi(p, j) \rhohat_{+}^{(j)}} \cdot p.
\end{align}
Under the locally greedy algorithm, the probability of successfully distinguishing between states $\rhohat_{+}$ and $\rhohat_{-}$ is given by 
\begin{align}
\label{eq:prob_success_lg}
P_{\mathrm{s, lg}}(q, \rhohat_{\pm}) \triangleq 1 - P_{\mathrm{err}, N} .
\end{align}

\subsubsection*{Plateau with locally greedy algorithm}

We observe the plateau in performance using the following experimental setup (dropping for now the prior $q$ as we assume $q = \frac{1}{2}$ in all cases unless specified otherwise):
\begin{enumerate}

\item Choose a set of allowed depolarizing parameters and number of trials.
In this case, we choose $\mathcal{S}_{\mathrm{dep}} = \{0.01, 0.05, 0.1, 0.3\}$ and $n_{\mathrm{trial}} = 1000$.

\item Generate $\theta_\pm^{(t)} \in (0, 2 \pi)$ uniformly, where $t \in [n_{\mathrm{trial}}]$ denotes the trial index. 

\item For each $\gamma \in \mathcal{S}_{\mathrm{dep}}$, define the corresponding qubit quantum states $\rhohat_{\pm}(\gamma, t) \triangleq (1-\gamma) \ketbra{\theta_\pm^{(t)}} + \frac{\gamma}{2} I$, where 
\begin{align}
\ket{\theta} \triangleq \cos\frac{\theta}{2} \ket{0} + \sin \frac{\theta}{2} \ket{1}.
\end{align}
Note that the subscript $\pm$ in $\theta_\pm^{(t)}$ is used to represent that the angles are chosen independently for the $\rhohat_{+}$ and $\rhohat_{-}$ states.

\item For all $\gamma \in \mathcal{S}_{\mathrm{dep}}$ and all $N = 1, 2, \ldots, 12$, we define the candidate QTPS generated by the random sampling as 
denoted by
\begin{align}
    P_{\mathrm{succ}}(N, \gamma) = \frac{1}{n_{\mathrm{trial}}} \sum_{t=1}^{n_{\mathrm{trial}}} P_{\mathrm{s,lg}} \left( {\rhohat_\pm(\gamma, t)}^{\otimes N} \right),
\end{align}

where $P_{\mathrm{s,lg}}(\rhohat_{\pm})$ is the success probability for the locally greedy algorithm and candidate states $\rhohat_\pm(\gamma, t)^{\otimes N}$.  \\
In the above, we randomly sample a set of pure states $\{ \ket{\theta_{\pm}^{(t)}} \bra{\theta_{\pm}^{(t)}} \}|_{t=1}^{n_{trial}}$ and generate the corresponding set of candidate states $\{ \rho_{\pm}(\gamma, t)^{\otimes N} \}|_{t=1}^{n_{trial}}$ for each $N$ and $\gamma$. Thus, 
$P_{\mathrm{succ}}(N, \gamma)$ represents the Monte Carlo average of performance for fixed $N$ and $\gamma$.

\end{enumerate}

\begin{figure}
 \centering
  \scalebox{0.8}{
\begin{tikzpicture}

\begin{axis}[
legend cell align={left},
legend style={at={(0.97,0.03)}, anchor=south east, draw=white!80.0!black},
tick align=outside,
tick pos=left,
x grid style={white!69.01960784313725!black},
xlabel={$N$},
xmin=1, xmax=12,
y grid style={white!69.01960784313725!black},
ylabel={$P_{\text{succ}}(N, \gamma)$},
ymin=0.6, ymax=1.0,
xmajorgrids,
ymajorgrids
]
\addplot [semithick, red, mark=+, mark options={solid}]
table [row sep=\\]{%
1	0.816100664673981 \\
2	0.861943635389522 \\
3	0.881259653516946 \\
4	0.898967518603093 \\
5	0.908306069241685 \\
6	0.914278660180836 \\
7	0.920185066740732 \\
8	0.926652831585514 \\
9	0.928159974664925 \\
10	0.931570282108423 \\
11	0.933287294167629 \\
12	0.937217910095904 \\
};
\addlegendentry{$\gamma = 0.01$}

\addplot [semithick, blue, mark=o, mark options={solid}]
table [row sep=\\]{%
1	0.803328919615036 \\
2	0.844179012424029 \\
3	0.865175394540013 \\
4	0.881412892319802 \\
5	0.889121342729621 \\
6	0.893699515930499 \\
7	0.898462837874322 \\
8	0.903363732670417 \\
9	0.903453421518272 \\
10	0.905597957767899 \\
11	0.906436497219879 \\
12	0.908888160489424 \\
};
\addlegendentry{$\gamma = 0.05$}

\addplot [semithick, black, mark=x, mark options={solid}]
table [row sep=\\]{%
1	0.787364239518308 \\
2	0.822303575798231 \\
3	0.844509281711572 \\
4	0.859170894142771 \\
5	0.865896987525865 \\
6	0.869148856073478 \\
7	0.872664991535815 \\
8	0.875835294504623 \\
9	0.874510628647767 \\
10	0.87598649679433 \\
11	0.875995992130248 \\
12	0.877499195709819 \\
};
\addlegendentry{$\gamma = 0.1$}

\addplot [semithick, green!50.0!black, mark=triangle, mark options={solid}]
table [row sep=\\]{%
1	0.723505519075038 \\
2	0.74063678311536 \\
3	0.77970051373649 \\
4	0.787692435350777 \\
5	0.797293897078216 \\
6	0.797754428716 \\
7	0.801694096037334 \\
8	0.803102698213465 \\
9	0.8007408774783 \\
10	0.801703427216124 \\
11	0.801600354385908 \\
12	0.802214548357316 \\
};
\addlegendentry{$\gamma = 0.3$}
\end{axis}

\end{tikzpicture}
  }
  \caption{\label{fig:plateaucopies} Comparison of probability of success for varying $\gamma$ in the case of identical copies, as a function of the number of available systems. Based on the computational results, we observe that as the depolarizing parameter increases, the probability of success levels off for large $N$.}
\end{figure}
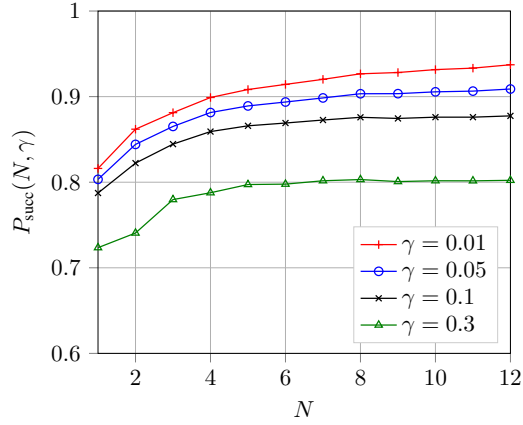

We plot the results of this computational experiment in Fig.~\ref{fig:plateaucopies}.
We observe that the average probability of success (asymptotically) approaches a value strictly less than $1$ when the depolarizing parameter is sufficiently high.
In the limiting case $\gamma \rightarrow 0$, the probability of success must approach $1$ with increasing $N$ because the locally greedy approach recovers the optimal Helstrom performance (see Theorem~\ref{thm:pure_states}).
Next, we prove a result that explains this plateau in performance and then define a modified locally greedy approach which overcomes this sub-optimality problem.
Note that an arbitrary qubit state (density matrix) can always be expressed as a pure state passed through a depolarizing channel, because this procedure can define any state in the Bloch sphere~\cite{Nielsen-2010}.

\begin{lemma}
\label{lem:depolarized_bd}
 Consider two $d$-dimensional qudit states $\rhohat_+$ and $\rhohat_-$. Suppose that we are given $\rhohat_+$ with probability $q$ and $\rhohat_-$ with probability $1-q$ where $q \leq \frac{1}{2}$. The depolarized versions of $\rhohat_{\pm}$ are defined as follows:
  \begin{align}
    \rhohat_\pm^{\mathrm{dep}} := (1 - \gamma) \rhohat_{\pm} + \frac{\gamma}{d} I.
  \end{align}
  Consider sufficiently small $\gamma$, such that $\frac{\gamma}{1 - \gamma} \frac{1 - 2q}{d}$ is less than the 
  magnitude of the largest negative eigenvalue of $(1-q) \rhohat_- - q \rhohat_+$. Then if the probability of distinguishing $\rhohat_+$ and $\rhohat_-$ is $P_{\mathrm{succ}}$, the probability of distinguishing $\rhohat_+^{\mathrm{dep}}$ and $\rhohat_-^{\mathrm{dep}}$ is given by
  \begin{align}
    P_{\mathrm{succ}}^{\mathrm{dep}} = \gamma q + \frac{\gamma(1 - 2q)k}{d}  + (1-\gamma) P_{\mathrm{succ}},
  \end{align}
  where $k$ is the rank of the Helstrom projector distinguishing $\rhohat_+$ and $\rhohat_-$.
\end{lemma}
\begin{IEEEproof}
  The Helstrom measurement is given by the orthogonal projector onto the positive eigenspace of the operator $\big[ (1-q) \rhohat_- - q \rhohat_+ \big]$. More explicitly, it is given by the orthogonal projector onto the vector space spanned by all eigenstates $\ket{v}$ such that
  \[
    \bra{v}\Big[ (1-q) \rhohat_- - q \rhohat_+ \Big] \ket{v} \geq 0.
  \]
  Let us denote this projector as $\Pi_{\mathrm{Hel}}$. Using this orthogonal projector, the probability of success is given by:
  \begin{align*}
    P_{\mathrm{succ}}  & = q  \Tr{(I - \Pi_{\mathrm{Hel}}) \rhohat_+} + (1-q) \Tr{\Pi_{\mathrm{Hel}} \rhohat_-} \\
    \addlinespace[1ex]
     & = q + \Tr{\Pi_{\mathrm{Hel}}\Big[ (1-q) \rhohat_- - q \rhohat_+ \Big]}.
  \end{align*}
  Now let us consider the optimal measurement for distinguishing $\rhohat_+^{\mathrm{dep}}$ and $\rhohat_-^{\mathrm{dep}}$. Calculating the analogous operator for $\rhohat_+^{\mathrm{dep}}$ and $\rhohat_-^{\mathrm{dep}}$ gives
  \begin{align*}
    (1-q) \rhohat_-^{\mathrm{dep}} - q \rhohat_+^{\mathrm{dep}} & = (1-q)\Big[ (1 - \gamma) \rhohat_{-} + \frac{\gamma}{d} I \Big] - q \Big[ (1 - \gamma) \rhohat_{+} + \frac{\gamma}{d} I \Big] \\
    \addlinespace[1ex]
                            & = \frac{\gamma(1-2q)}{d} I + (1 - \gamma) \Big[ (1-q) \rhohat_- - q \rhohat_+ \Big].
  \end{align*}
  Since $\gamma < 1$, we can divide by $1 - \gamma$ without changing the positive eigenspace. Therefore, the Helstrom optimal measurement projects onto the space of eigenstates $\ket{v}$ such that the following is positive:
\begin{align}
\label{eq:eigenvalue_shift}
    \bra{v} \bigg( \frac{\gamma}{1-\gamma}\,\frac{1-2q}{d} I + \Big[ (1-q) \rhohat_- - q \rhohat_+ \Big]\bigg) \ket{v} = \bra{v}\Big[ (1-q) \rhohat_- - q \rhohat_+ \Big] \ket{v} + \frac{\gamma}{1-\gamma}\,\frac{1-2q}{d}.
  \end{align}
  Therefore, if $\gamma$ is sufficiently small, then the optimal projector distinguishing $\rhohat_+^{\mathrm{dep}}$ and $\rhohat_-^{\mathrm{dep}}$ is $\Pi_{\mathrm{Hel}}$. Hence, for $\gamma$ sufficiently small, we have:
  \begin{align*}
    P_{\mathrm{succ}}^{\mathrm{dep}} & = q + \Tr{\Pi_{\mathrm{Hel}}\Big[ (1-q) \rhohat_-^{\mathrm{dep}} - q \rhohat_+^{\mathrm{dep}} \Big]} \\
                        & = q + \frac{\gamma(1-2q)}{d} \Tr{\Pi_{\mathrm{Hel}}} + (1 - \gamma)\Tr{\Pi_{\mathrm{Hel}} \Big[ (1-q) \rhohat_- - q \rhohat_+ \Big]} \\
                        & = \gamma q + \frac{\gamma(1-2q)k}{d} + (1- \gamma) P_{\mathrm{succ}}. \tag*{\IEEEQEDhere}
\end{align*}
\end{IEEEproof}

\noindent In the case of qubits, this lemma implies the following corollary.

\begin{corollary}
\label{cor:succ_prob_bound}
Consider the problem of distinguishing between two distinct single qubit states $\rhohat_+^{\textrm{dep}}$ and $\rhohat_-^{\textrm{dep}}$ with prior probabilities $q$ and $1-q$ respectively. 
Assume that $\rhohat_+^{\mathrm{dep}}$ and $\rhohat_-^{\mathrm{dep}}$ are depolarized such that there exist pure states $\ketbra{\psi_+}{\psi_+}$, $\ketbra{\psi_-}{\psi_-}$ such that 
\[ \gamma_{\pm } \in [0, 1] \ \ \text{and} \ \ 
    \rhohat_\pm^{\mathrm{dep}} \triangleq (1 - \gamma_{\pm}) \ketbra{\psi_\pm}{\psi_\pm} + \frac{\gamma_{\pm}}{2} I.
\]
For any choice of $\gamma_{\pm}, \  q \in [0, 1]$ the probability of correctly distinguishing $\rhohat_+^{\mathrm{dep}}$ and $\rhohat_-^{\mathrm{dep}}$, $P_{\mathrm{succ}}^{\mathrm{dep}}$ satisfies
\begin{align}
P_{\mathrm{succ}}^{\mathrm{dep}} \leq \max\left\{ 1-q, q, 1 - \frac{\gamma_{min}}{2}\right\}
\end{align}
where $\gamma_{min} \triangleq \text{min}(\gamma_{+}, \gamma_{-})$.
\end{corollary}

\begin{IEEEproof}
Let us denote the Helstrom measurement for $\{\ket{\psi_{+}}\bra{\psi_{+}}, \ket{\psi_{-}}\bra{\psi_{-}} \}$ by $\Pi_{\mathrm{Hel}, \ket{\psi_{\pm}}\bra{\psi_{\pm}}}$ and the Helstrom measurement for $\{\rho_{+}^{\textrm{dep}}, \rho_{-}^{\textrm{dep}} \}$ by $\Pi_{\mathrm{Hel}, \rho_{\pm}}$. Since $\rho_{\pm}^{\textrm{dep}}$ are qubit states, $\text{rank}(\Pi_{\mathrm{Hel}, \rho_{\pm}})$ is 0, 1, or 2.\\

If $\text{rank}(\Pi_{\mathrm{Hel},\rhohat_\pm}) =0$, then $\Pi_{\mathrm{Hel},\rhohat_\pm}=0$ and
\begin{align*}
P_{\mathrm{succ}}^{\mathrm{dep}} &=
 q + \trays{\Pi_{\mathrm{Hel},\rhohat_\pm} \Big( (1-q) \rhohat_-^{\mathrm{dep}} - q \rhohat_+^{\mathrm{dep}} \Big)} = q.
\end{align*}

If $\text{rank}(\Pi_{\mathrm{Hel},\rhohat_\pm}) =2$, then $\Pi_{\mathrm{Hel},\rhohat_\pm}=I$ and
\begin{align*}
P_{\mathrm{succ}}^{\mathrm{dep}} &=
 q + \trays{\Pi_{\mathrm{Hel},\rhohat_\pm} \Big( (1-q) \rhohat_-^{\mathrm{dep}} - q \rhohat_+^{\mathrm{dep}} \Big)} = 1-q.
\end{align*}
 


  
Finally, consider the case where $\text{rank}(\Pi_{\mathrm{Hel},\rhohat_\pm}) =1$. The state discrimination problem between $\{ \rho_{+}^{\mathrm{dep}}, \rho_{-}^{\mathrm{dep}} \}$ is physically equivalent to a black box which outputs one of the following four separate discrimination problems:
\begin{align*}
\Big\{ \{ \ket{\psi_{+}}\bra{\psi_{+}}, \ket{\psi_{-}}\bra{\psi_{-}} \}, \{ \ket{\psi_{+}}\bra{\psi_{+}}, \frac{\mathbb{I}}{2} \},  \{ \frac{\mathbb{I}}{2}, \ket{\psi_{-}}\bra{\psi_{-}} \}, \{ \frac{\mathbb{I}}{2}, \frac{\mathbb{I}}{2} \} \Big\},
\end{align*}
with probabilities 
\begin{align*}
    \Big\{ p_{1}, p_{2}, p_{3}, p_{4} \Big\} \triangleq \Big\{ (1 - \gamma_{+})(1-\gamma_{-}), (1- \gamma_{+})\gamma_{-}, \gamma_{+}(1-\gamma_{-}), \gamma_{+}\gamma_{-} \Big\} 
\end{align*}
respectively. (This follows from viewing $\rho_{\pm}^{\mathrm{dep}}$ as corresponding to a quantum system prepared in state $\ket{\psi_{\pm}}\bra{\psi_{\pm}}$ with probability $1 - \gamma_{\pm}$ and prepared in state $\frac{\mathbb{I}}{2}$ with probability $\gamma_{\pm}$.) \\
We denote by $P_{\mathrm{succ}}(\rho_{+}, \rho_{-}, \Pi)$ the probability of successfully discriminating between $\{ \rho_{+}, \rho_{-}\}$ given measurement $\{ \Pi, \mathbb{I} - \Pi \}$ where the prior is implicitly defined as $q$. Then we can upper bound the success probability as
\begin{align*}
    P_{\mathrm{succ}}^{\mathrm{dep}} \leq & p_{1} \max_{\ket{\psi_{+}}, \ket{\psi_{-}}, \Pi } P_{\mathrm{succ}} \Big(\ket{\psi_{+}} \bra{\psi_{+}}, \ \ket{\psi_{-}}\bra{\psi_{-}}, \ \Pi \Big)  + p_{2} \max_{\ket{\psi_{+}}, \Pi } P_{\mathrm{succ}}\Big(\ket{\psi_{+}} \bra{\psi_{+}}, \ \frac{\mathbb{I}}{2}, \ \Pi\Big)\\
    & +  p_{3} \max_{\ket{\psi_{-}}, \Pi } P_{\mathrm{succ}} \Big( \frac{\mathbb{I}}{2}, \ \ket{\psi_{-}} \bra{\psi_{-}}, \ \Pi \Big) 
     + p_{4} \times \frac{1}{2} \\
    = & p_{1} P_{\mathrm{succ}}\Big(\ket{0} \bra{0}, \ \ket{1}\bra{1}, \ \Pi_{\mathrm{Hel}, \{\ket{0}\bra{0}, \ket{1}\bra{1}\} } \Big)  + p_{2} P_{\mathrm{succ}}\Big( \ket{0} \bra{0}, \  \frac{\mathbb{I}}{2}, \ \Pi_{\mathrm{Hel}, \{\ket{0}\bra{0}, \ket{1}\bra{1}\}} \Big) \\
    & +  p_{3} P_{\mathrm{succ}} \Big( \frac{\mathbb{I}}{2}, \ \ket{1} \bra{1}, \ \Pi_{\mathrm{Hel}, \{\ket{0}\bra{0}, \ket{1}\bra{1}\}}\Big) + \frac{p_{4}}{2}  \\
    = & P_{\mathrm{succ}}\Big((1-\gamma_{+})\ket{0}\bra{0} + \frac{\gamma_{+}}{2}\mathbb{I}, \  (1-\gamma_{-})\ket{1}\bra{1} + \frac{\gamma_{-}}{2}\mathbb{I}, \  \Pi_{\mathrm{Hel}, \{\ket{0}\bra{0}, \ket{1}\bra{1}\}} \Big).
\end{align*}
Thus, the success probability for $\rho_{\pm}^{dep}$ is upper bounded by the success probability 
when $\ket{\psi_{+}}$ and $\ket{\psi_{-}}$ are orthogonal (w.l.o.g. we have set $\ket{\psi_{+}} = \ket{0}$ and $\ket{\psi_{-}} = \ket{1}$). Upon solving for $P_{\mathrm{succ}}\Big(\ket{0}\bra{0}^{\mathrm{dep}}, \ \ket{1}\bra{1}^{\mathrm{dep}}, \  \Pi_{\mathrm{Hel}, \{\ket{0}\bra{0}, \ket{1}\bra{1}\}} \Big)$, it immediately follows that: 
\begin{align*}
   P_{\mathrm{succ}}^{\mathrm{dep}} \leq  (1- \frac{\gamma_{+}}{2})q + (1 - \frac{\gamma_{-}}{2})(1-q) \leq 1 - \frac{\gamma_{min}}{2}.
\end{align*}

 \IEEEQEDhere
\end{IEEEproof}

Assuming w.l.o.g. that $q \leq \frac{1}{2}$, observe that $1 - q \geq 1 - \frac{\gamma}{2}$ implies $\gamma \geq 2q$ and therefore $\frac{\gamma}{1 - \gamma} \frac{(1 - 2q)}{qd} \geq 1$ ($d = 2$).
In the notation of Lemma~\ref{lem:depolarized_bd}, set $\rhohat_{\pm} = \ketbra{\psi_{\pm}}$.
Since the spectrum of $\big[ (1-q) \ketbra{\psi_-} - q \ketbra{\psi_+} \big]$ lies in the interval $[-1, 1]$, eq.~\eqref{eq:eigenvalue_shift} implies that the smallest eigenvalue of $\big[ (1-q) \rhohat_-^{\mathrm{dep}} - q \rhohat_+^{\mathrm{dep}} \big]$ will now be non-negative and hence $\Pi_{\mathrm{Hel},\rhohat_\pm} = I$ will be trivial.
Hence, in this scenario, the Helstrom measurement is equivalent to guessing according to the prior. 

In summary, this corollary implies that, for equally depolarized states, once the prior is updated so that either $q$ or $1-q$ is greater than $1 - \frac{\gamma}{2}$, the locally greedy algorithm will be stuck making trivial measurements for all subsequent subsystems and therefore the error will not approach $0$ as $N \rightarrow \infty$. In Appendix~\ref{sec:additional_lgm} we show that the locally greedy method also exhibits plateaus in more general scenarios.

This result provides motivation for us to modify the conventional locally greedy method discussed above (first introduced by~\cite{Acin-physreva05} and~\cite{Higgins-physreva11}). In particular, a ``modified Helstrom'' measurement is implemented whenever the Helstrom measurement is trivial (namely, $\Pi_{\mathrm{Hel}, \rho_{\pm}} \in \{I, 0\}$). In the next section, we introduce this modified locally greedy method (MLG method) and show that for arbitrary, qubit-subsystem $\rho_{\pm}$, we have $P_{s, mlg}(\rho_{\pm}) \geq P_{s, lg}(\rho_{\pm})$ where $P_{s, mlg}(\rho_{\pm})$ is the probability of successful discrimination under the MLG method. We further show that for any $\rho_{\pm}$, $P_{s, mlg}(\rho_{\pm}) \rightarrow 1$ as the number of subsystems $j$ such that $\rho_{+}^{(j)} \ne \rho_{-}^{(j)}$ approaches infinity.

\subsection{Modified Locally Greedy (MLG) algorithm}

Like the locally greedy algorithm, the MLG algorithm 
updates the prior after each measurement round. Then, it performs the modified Helstrom measurement according to the new prior, with the modified Helstrom measurement defined as:
\begin{align*}
    \Pi^{*}(p, j) \triangleq \begin{cases}
    \Pi(p, j) & \text{if} \   \Pi(p, j) \notin \{\mathbb{I}, 0\} \\
    \ket{v_{\lambda_{\mathrm{max}}}}\bra{v_{\lambda_{\mathrm{max}}}} & \text{if} \  \Pi(p, j) = 0, \ \text{where} \displaystyle \ \lambda_{\mathrm{max}} \triangleq \max_{\lambda} \Big\{ \lambda \ \Big| \ \big( (1-p)\rho_{-}^{(j)} - p \rho_{+}^{(j)} \big) \ket{v_\lambda} = \lambda \ket{v_{\lambda}} \Big\} \\
    \mathbb{I} - \ket{v_{\lambda_{\mathrm{min}}}}\bra{v_{\lambda_{\mathrm{min}}}} & \text{if} \  \Pi(p, j) = \mathbb{I}, \ \text{where} \displaystyle \ \lambda_{\mathrm{min}} \triangleq \min_{\lambda} \Big\{ \lambda \ \Big| \ \big( (1-p)\rho_{-}^{(j)} - p \rho_{+}^{(j)} \big) \ket{v_\lambda} = \lambda \ket{v_{\lambda}} \Big\}
    \end{cases}
\end{align*}
where the final state is decoded as $\hat{\rho} = \rho_{+}$ if $C_{N}^{\sigma}(q, \mathbf{a}_{[N]}^{\sigma}, \mathbf{d}_{[N]}^{\sigma} ) \geq \frac{1}{2}$ and as $\hat{\rho} = \rho_{-}$ otherwise. 

Whenever the Helstrom measurement is nontrivial, the modified Helstrom measurement is equivalent to the Helstrom and thus locally optimal by definition. In the case where the Helstrom measurement is trivial, then any other measurement and outcome would lead to identical posterior-based decoding (i.e. any measurement is locally optimal). The modified Helstrom measurement takes advantage of this degeneracy to replace the trivial Helstrom measurement with a more informative measurement. 

Consider the measurement given by the set of projectors $\Big\{ \ket{v_{\lambda}}\bra{v_{\lambda}} \ \Big| \ \big( (1-p)\rho_{-}^{(j)} - p \rho_{+}^{(j)} \big) \ket{v_\lambda} = \lambda \ket{v_{\lambda}} \Big\}$ and w.l.o.g. let $p \geq \frac{1}{2}$. Given a measurement outcome corresponding to projector $\ket{v_{\lambda}}\bra{v_{\lambda}}$, the posterior-based decoding is uniquely determined by the sign of $\lambda$, with larger values of $\lambda$ being a stronger predictor that $\rho = \rho_{-}$. 

The Helstrom measurement then partitions these projectors by the sign of their eigenvalues and groups all projectors together into a trivial measurement when all eigenvalues have the same sign. In the case where the Helstrom measurement is trivial, the modified Helstrom measurement instead partitions the projectors based on the ordering of their eigenvalues. Thus, it separates out the projector most strongly predictive of the less-likely candidate state.  


\begin{lemma}
\label{lem:helstrom_nonunique}
Let $\rho_{\pm}^{(j)}$ and $p$ be such that $\Pi(p, j) = \mathbb{I}$ or $\Pi(p, j) = 0$. Then 
$\max_{\Pi} \Big( \frac{p \mathrm{Tr}[\rho_{+}^{(j)} \Pi]}{\mathrm{Tr}[\Pi ( (1-p) \rho_{-}^{(j)} + p \rho_{+}^{(j)}]} \Big) < \frac{1}{2}$ or \\ $\min_{\Pi} \Big( \frac{p \mathrm{Tr}[\rho_{+}^{(j)} \Pi]}{\mathrm{Tr}[\Pi ( (1-p) \rho_{-}^{(j)} + p \rho_{+}^{(j)}]} \Big)  \geq \frac{1}{2}$ respectively. Namely, any local measurement is optimal given posterior-based decoding.
\end{lemma}
\begin{IEEEproof} Define $M \triangleq (1-p) \rho_{-} - p \rho_{+}$ and let the resulting projector be $\Pi_{h}(p, \rho_{\pm}) = \mathbb{I}$. Then the eigenvalues of $M$ satisfy $\lambda_{j} > 0$ \ $\forall j$, and as $M$ is Hermitian the eigenvectors $\{ \ket{v_{j}} \}$ are orthogonal and form a basis. Any projector diagonal in this basis may be defined $\Pi_{S} \equiv \sum_{j \in S} \ket{v_{j}}\bra{v_{j}}$ for some set of indices $S$. It follows that $\text{Tr}[M \Pi_{s}] = \sum_{j \in S} \lambda_{j} > 0$, so $\text{Tr}[ \Pi_{S} p \rho_{+}] < \text{Tr} [\Pi_{S} (1-p) \rho_{-}]$. Then, the updated prior upon obtaining measurement corresponding to $\Pi_{S}$ is:
\begin{align*}
p' = \frac{\text{Tr}[\Pi_{S} \rho_{+} p]}{\text{Tr}[\Pi_{S}p \rho_{+}] + \text{Tr}[\Pi_{S} (1-p) \rho_{-}]} < \frac{1}{2} \ \ \ \ \ \ \forall S
\end{align*} 
Now suppose the projector is diagonal in an arbitrary basis $\{ \ket{w_{k}} \}$ s.t. $\ket{w_{k}} = \sum_{j} \alpha_{k, j} \ket{v_{j}}$ where $\{ \alpha_{k, j} \}$ form the entries of some unitary operator. Then it is sufficient to show that $\text{Tr}[M \ket{w_{k}} \bra{w_{k}}] > 0$ for all $ k$, since then $\text{Tr} [M \sum_{k \in S} \ket{w_{k}} \bra{w_{k}}]> 0$ for all $S$. We observe:
\begin{align*}
\text{Tr}[M \ket{w_{k}} \bra{w_{k}}] &= \sum_{j, j'}  \alpha_{k, j} \alpha_{k, j'}^{*} \lambda_{j} \text{Tr} [\ket{v_{j}} \bra{v_{j'}}] \\
&= \sum_{j}  |\alpha_{k, j}|^{2} \lambda_{j} >0
\end{align*}
Similarly, for any basis $\{\ket{w_{k}} \}$, then $\text{Tr}[M \ket{w_{k}} \bra{w_{k}}] \leq 0$ if $\Pi_{h}(p, \rho_{\pm}) = 0$.
\end{IEEEproof}

Denote by $P_{\mathrm{s, mlg}}(q, \rho_{\pm})$ the success probability of distinguishing $\{\rho_{+}, \rho_{-}\}$ with initial prior $q$ using the MLG algorithm. We now show that the MLG method exhibits the desired asymptotic behaviour in the limit of large $N$. Additionally, we show $P_{s,mlg}( \rho_{\pm}) \geq P_{s,lg}(\rho_{\pm})$ for all $\rho_{\pm}$ so the MLG algorithm always performs at least as well as the LG algorithm. 

\begin{corollary}
\label{cor:MLG_asymptotic}
For any $\rho_{\pm}$ where $\rho_{+}^{(j)} \ne \rho_{-}^{(j)}$ for all subsystems $j$, then in the limit $N \rightarrow \infty$, $P_{s, mlg}(q, \rho_{\pm}) = 1$.
\end{corollary}
\begin{IEEEproof}
It is sufficient to show that for all $j \in \{0, 1, ..., N\}$ and for all $p_{j} \in (0, 1)$ we have

\begin{align*}
f_{+}(p_{j}) \triangleq \mathbb{E}[p_{j+1}| \rho = \rho_{+}^{(j)}, \Pi^{*}(p_{j}, j)] &> p_{j} \ \text{and} \   f_{-}(p_{j}) \triangleq \mathbb{E}[p_{j+1}| \rho = \rho_{-}^{(j)}, \Pi^{*}(p_{j}, j)] < p_{j}, \\
\end{align*}
and that $f_{\pm}(p_{j})$ is continuous with no fixed points other than $p_{j} = 0$ or $1$. For simplicity, we drop the superscript on $\rho_{\pm}^{(j)}$ in the following whenever the subsystem index is unambiguous. We denote the modified Helstrom measurement as $\Pi = \Pi^{*}(p_{j}, j)$, such that by definition $\text{Tr}[\Pi \rho_{-}] > \text{Tr}[\Pi \rho_{+}]$. \\
Let $x \triangleq \text{Tr}[\Pi \rho_{-}] - \text{Tr}[\Pi \rho_{+}] = \text{Tr}[\Pi^{\perp} \rho_{+}] - \text{Tr}[\Pi^{\perp} \rho_{-}] $ s.t. $x \in (0, 1]$. Then, there exists $y \in [\frac{x}{2}, 1-\frac{x}{2}]$ such that the conditional measurement probabilities may be represented as follows:
\begin{align*}
\text{Tr}[\Pi^{\perp} \rho_{\pm}] &= y \pm \frac{x}{2}, \ \
\text{Tr}[\Pi \rho_{\pm}] = 1 - y \mp \frac{x}{2}
\end{align*}

Finally, we calculate $f_{+}(p_{j})$ as follows:
\begin{align*}
& f_{+}(p_{j}) = \text{Tr}[\Pi \rho_{+}] \Big( \frac{p_{j} \text{Tr}[\Pi \rho_{+}]}{\text{Tr}[\Pi(p_{j} \rho_{+} + (1 - p_{j}) \rho_{-})]} \Big) +  \text{Tr}[\Pi^{\perp} \rho_{+}] \Big( \frac{p_{j} \text{Tr}[\Pi^{\perp} \rho_{+}]}{\text{Tr}[\Pi^{\perp}(p_{j} \rho_{+} + (1 - p_{j}) \rho_{-})]} \Big) \\
& = p_{j} \Bigg(\frac{(1 - y - \frac{x}{2})^{2}}{p_{j} (1 - y - \frac{x}{2}) + (1-p_{j})(1-y+ \frac{x}{2})} + \frac{(y + \frac{x}{2})^{2}}{p_{j} (y + \frac{x}{2}) + (1 - p_{j})(y - \frac{x}{2})}  \Bigg) \\
& > p_{j}
\end{align*}
where the final line follows from solving symbolically for the range $p_{j} \in (0, 1); \ x \in (0, 1); \ y \in [\frac{x}{2}, 1 - \frac{x}{2}]$.  We then check for any fixed points $p_{j}^{*} = f_{+}(p_{j}) $. This results in the condition $p_{j}^{*} (-x^{2} + 2 p_{j}^{*} x^{2} - (p_{j}^{*} x)^{2}) = 0$ so the only fixed points are $p_{j}^{*} = 0$ or $1$. Additionally, it is clear that $f_{+}(p_{j})$ is continuous in $p_{j}$.


A similar argument holds for the case $\rho = \rho_{-}$. Thus, the probability of success converges to $1$ under the MLG algorithm.
\end{IEEEproof}

\begin{figure}
 \centering
  \scalebox{0.8}{
 \begin{tikzpicture}

\begin{axis}[
legend cell align={left},
legend entries={{$\gamma=0.01$},{$\gamma=0.05$},{$\gamma=0.1$},{$\gamma=0.3$}},
legend style={at={(0.97,0.03)}, anchor=south east, draw=white!80.0!black},
tick align=outside,
tick pos=left,
x grid style={white!69.01960784313725!black},
xlabel={N},
xmajorgrids,
xmin=1, xmax=12,
y grid style={white!69.01960784313725!black},
ylabel={$P_{\text{succ}}(N, \gamma)$},
ymajorgrids,
ymin=0.7, ymax=1.01
]
\addlegendimage{no markers, red}
\addlegendimage{no markers, blue}
\addlegendimage{no markers, black}
\addlegendimage{no markers, green!50.0!black}
\addplot [semithick, red, mark=+, mark size=3, mark options={solid}]
table [row sep=\\]{%
1	0.821328726475527 \\
2	0.917971731260055 \\
3	0.960063868873501 \\
4	0.979381784957372 \\
5	0.988251813438165 \\
6	0.993999216728556 \\
7	0.996400561902347 \\
8	0.997791891135667 \\
9	0.99866045226735 \\
10	0.999247641489843 \\
11	0.99954209083421 \\
12	0.999734433603017 \\
};
\addplot [semithick, blue, mark=*, mark size=1.5, mark options={solid}]
table [row sep=\\]{%
1	0.808345746840902 \\
2	0.900071278319693 \\
3	0.942640105565071 \\
4	0.964929452498443 \\
5	0.977135556907065 \\
6	0.985778556648899 \\
7	0.990442234276101 \\
8	0.993578264859847 \\
9	0.995695013775666 \\
10	0.997213690088247 \\
11	0.998136177222562 \\
12	0.998763589147328 \\
};
\addplot [semithick, black, mark=x, mark size=3, mark options={solid}]
table [row sep=\\]{%
1	0.792117023336085 \\
2	0.877869678197398 \\
3	0.921561074954545 \\
4	0.947309854343065 \\
5	0.963041655432957 \\
6	0.974642484186956 \\
7	0.981748894034004 \\
8	0.986910519524632 \\
9	0.99062079550542 \\
10	0.993403409777118 \\
11	0.995261002767565 \\
12	0.996592966040238 \\
};
\addplot [semithick, green!50.0!black, mark=triangle*, mark size=3, mark options={solid}]
table [row sep=\\]{%
1	0.727202128546003 \\
2	0.790858537640391 \\
3	0.836326426026325 \\
4	0.869249677074009 \\
5	0.893297447084324 \\
6	0.912853459882475 \\
7	0.927496576336425 \\
8	0.939733588687299 \\
9	0.949750089163797 \\
10	0.958197555763091 \\
11	0.964832477158153 \\
12	0.970352640625428 \\
};
\end{axis}

\end{tikzpicture}
  }
  \caption{\label{fig:MLGcopies} Comparison of probability of success for varying $\gamma$ in the case of identical copies, as a function of the number of available systems using the MLG algorithm. We observe that as the depolarizing parameter increases, the probability of success no longer levels off for large $N$.}
\end{figure}
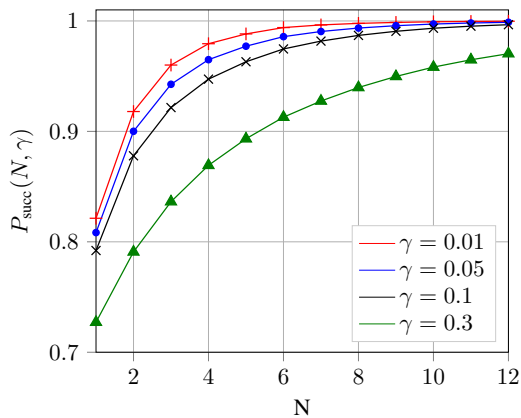

From the above, we can conclude that $P_{s, mlg}(\rho_{\pm}) \geq P_{s, lg}(\rho_{\pm})$ as the MLG and locally greedy methods are equivalent whenever the Helstrom measurement is nontrivial. When the Helstrom measurement is trivial, it follows that the MLG method does strictly better. The improved asypmptotic behaviour of the MLG algorithm is depicted in Fig.~\ref{fig:MLGcopies}, where we repeat the previous experimental setup with the MLG algorithm, and plot the resulting $P_{succ}(N, \gamma) = \frac{1}{n_{trial}}\sum_{t=1}^{n_{trial}} P_{s, lg}(\rho_{\pm}(\gamma, t)^{\otimes N})$.

In the next section, we generalize to a dynamic programming based algorithm capable of optimizing over the order of subsystem measurement as well as the measurement performed on each subsystem.  

\section{Dynamic-Programming Algorithms}
\label{sec:dp_algorithms}

\subsection{Order-Optimized MLG Algorithm}

In comparison to the locally greedy algorithm, the main difference of the order is that one can choose $\sigma(j)$ carefully at each round $j$. To do this, we recursively compute an \emph{expected future risk} function $R_S \colon [0,1] \rightarrow [0,1]$, where $S$ denotes the set of subsystem indices that are yet to be measured and the domain corresponds to the current updated prior (CSP).
Formally, at round $j$,
\begin{align}
S \triangleq [N] \setminus \sigma([j-1]).
\end{align}
\begin{itemize}

\item For the base case $S = \emptyset$, one can make a hard decision on $C_N^\sigma (q, \mathbf{a}_{[N]}^\sigma, \mathbf{d}_{[N]}^\sigma ) $, i.e., by comparing it to $0.5$. 
Hence
\begin{align}
R_\emptyset \left( C_N^\sigma \left( q, \mathbf{a}_{[N]}^\sigma, \mathbf{d}_{[N]}^\sigma \right) \right) = \min \left( C_N^\sigma \left( q, \mathbf{a}_{[N]}^\sigma, \mathbf{d}_{[N]}^\sigma \right),\ 1 - C_N^\sigma \left( q, \mathbf{a}_{[N]}^\sigma, \mathbf{d}_{[N]}^\sigma \right) \right),
\end{align}
which can be written as function of $p \in [0, 1]$ as $R_\emptyset(p) = \min(p, 1 - p)$.

\item For the general case $S \neq \emptyset$ and $j = N - |S| + 1$, $N - |S|$ measurements have been performed. The goal is to choose the best subsystem $\sigma(j)$ to be measured next in order to minimize the expected error probability over the remaining measurements, i.e., 
\begin{align}
& R_S \left( C_{N - \abs{S}}^{\sigma} \left( q, \mathbf{a}_{[N-\abs{S}]}^{\sigma}, \mathbf{d}_{[N-\abs{S}]}^{\sigma} \right) \right) \nonumber \\
       & = \min_{k \in S} \sum_{d_k \in \mathcal{D}} \mathbb{P} \left( d_k \biggr\vert q, (\mathbf{a}_{[N -\abs{S}]}^{\sigma}, \mathbf{a}_k), \mathbf{d}_{[N - \abs{S}]}^{\sigma} \right) \cdot R_{S \setminus \{ k \}} \left( C_{N - \abs{S} + 1}^\sigma \left( q, (\mathbf{a}_{[N - \abs{S}]}^{\sigma}, \mathbf{a}_k), (\mathbf{d}_{[N - \abs{S}]}^{\sigma}, d_k) \right) \right),
\end{align}

where $\mathbf{a}_k = \Pi^{*}(p ,k)$ is the modified Helstrom measurement, with $p = C_{N - \abs{S}}^{\sigma} ( q, \mathbf{a}_{[N-\abs{S}]}^{\sigma}, \mathbf{d}_{[N-\abs{S}]}^{\sigma} ) = C_{j-1}^{\sigma} ( q, \mathbf{a}_{[j-1]}^{\sigma}, \mathbf{d}_{[j-1]}^{\sigma} )$.

This expression can be written as a function of $p \in [0, 1]$ as
\begin{align}
R_S(p) = \min_{k \in S} \sum_{d_k \in \mathcal{D}} \mathcal{L}(p, \Pi^{*}(p,k), d_k) \cdot R_{S \setminus \{ k \}} \left( \mathcal{P}(p, \Pi^{*}(p,k), d_k) \right). \label{bottom_up_recursion_order}
\end{align}
Hence, during the execution of the algorithm, the next mapping for $\sigma$ at round $j$ can be defined as
\begin{align}
\sigma(j) \triangleq \underset{k \in S}{\text{argmin}} \sum_{d_k \in \mathcal{D}} \mathcal{L}(p, \Pi^{*}(p,k), d_k) \cdot R_{S \setminus \{ k \}} \left( \mathcal{P}(p, \Pi^{*}(p,k), d_k) \right),
\end{align}
where $\mathcal{L}$ and $\mathcal{P}$ are defined in (11).
\end{itemize}

Similar to the case of identical copies, the measurement outcome probabilities are given by~\eqref{eq:outcome_prob_j} with $\Pi(p, j)$ replaced by $\Pi^{*}(p, j)$, the probability of error at round $j$ is given by~\eqref{eq:prob_error_j}, and the overall probability of success of the order-optimized locally greedy algorithm is given by~\eqref{eq:prob_success_lg} with $N$ replaced by $\sigma(N)$.

\subsection{Measurement- and Order-Optimized DYnamic (MOODY) Algorithm}
\label{sec:dynamic_program}

The MOODY algorithm is a generalization of the order-optimized locally greedy algorithm described above for the distinct subsystems scenario.
During execution of round $j$, the algorithm optimizes over all choices of $\sigma(j)$ as well as the measurement actions that could be performed over the chosen subsystem $\sigma(j)$.
Hence, the expected future risk function is given by
\begin{align}
R_S(p) = \min_{(k, \mathbf{a}_k) \in S \times \mathcal{A}} \sum_{d_k \in \mathcal{D}(\mathcal{A})} \mathcal{L}(p, \mathbf{a}_k, d_k) \cdot R_{S \setminus \{ k \}} \left( \mathcal{P}(p, \mathbf{a}_k, d_k) \right).
\end{align}
An optimal choice for the next subsystem, $k \in S$, and the optimal action to be performed on that subsystem, $\mathbf{a}_k \in \mathcal{A}$, are given by the minimizer $\hat{A}_S(p) = (k, \mathbf{a}_k)$ of the above function:
\begin{equation}
\hat{A}_S(p) \triangleq \underset{(k, \mathbf{a}_k) \in S \times \mathcal{A}}{\text{argmin}} \sum_{d_k \in \mathcal{D}(\mathcal{A})} \mathcal{L}(p, \mathbf{a}_k, d_k) \cdot R_{S \setminus \{ k \}} \left( \mathcal{P}(p, \mathbf{a}_k, d_k) \right).
\end{equation}
Therefore, during round $j$ of the algorithm, we have $j = N - |S| + 1$ and we choose 
\begin{align}
(\sigma(j), \mathbf{a}_{\sigma(j)}) = \hat{A}_S\left( C_{j-1}^{\sigma} \left( q, \mathbf{a}_{[j-1]}^{\sigma}, \mathbf{d}_{[j-1]}^{\sigma} \right) \right).
\end{align}

The MOODY algorithm can be summarized as below.
Once the DP subroutine is completed, we have a set of expected future error functions $ \clbrsv{ R_S }{ S \subseteq [N] } $ and a set of best measurement action functions $ \clbrsv{ A_S }{ S \subseteq [N] } $. 
Setting $p_{0} = \mathbb{P}(\hat{\rho}=\hat{\rho}_+) = q$ and  $ S_0 = \clbrs{ 1, \ldots, N } $, we then have for $ i = 0, \ldots, N-1 $:
\begin{subequations}
\begin{IEEEeqnarray}{rCl}
    P_i^{\mathrm{err}} &=& R_{S_i}(p_i), \\
    (\sigma(i+1), \mathbf{a}_{\sigma(i+1)}) &=& \hat{A}_{S_i}(p_i), \\
    \mathbb{P}(d_{\sigma(i+1)} = d \mid p_i, \mathbf{a}_{\sigma(i+1)}) &=& \mathcal{L}(p_i, \mathbf{a}_{\sigma(i+1)}, d), \\
    p_{i+1} &=& \mathcal{P}(p_i, \mathbf{a}_{\sigma(i+1)}, d_{\sigma(i+1)}), \\
    S_{i+1} &=& S_i \setminus \clbrs{\sigma(i+1)}.
\end{IEEEeqnarray}
\end{subequations}
Finally, after $ N $ rounds of measurements one makes the decision to decode $\rho$ as $\hat{\rho}$ for:
\begin{align}
    \hat{\rho} = 
    \begin{cases}
        \rhohat_{+}, &\text{ if } p_N > 0.5, \\
        \rhohat_{-}, &\text{ if } p_N < 0.5, \\
        \text{random guess}, &\text{ if } p_N = 0.5,\\
    \end{cases} \ \ \text{and}\ \ 
    P_{\mathrm{succ}} = \max( p_N, 1 - p_N ).
\end{align}

\subsection*{Implementation and Complexity}

We compute the functions $ R_S $ and $ \hat{A}_S $ using dynamic programming (DP), which requires mapping the states $ (q, \mathbf{a}_{[N-\abs{S}]}^\sigma, \mathbf{d}_{[N-\abs{S}]}^\sigma) \in [0,1] \times \mathcal{A}^{N-\abs{S}} \times \mathcal{D}^{N-\abs{S}} $ to $ C_{N-\abs{S}}^\sigma \rdbrs{ q, \mathbf{a}_{[N-\abs{S}]}^\sigma, \mathbf{d}_{[N-\abs{S}]}^\sigma } \in [0, 1] $.
Moreover, these functions can be stored for later use in problems with same states but different initial priors or if one needs to run MOODY on a larger system where the current system is a subsystem of it.

Since the interval $[0, 1]$ and action space $\mathcal{A}$ are not discrete, they are quantized to give a tractable implementation.
The computational complexity and memory requirements of DP are highly dependent on this quantization.
In our implementation, we again apply a quantized version of the above DP where the input $p$ is quantized into $Q_{p}$ equi-spaced points over $[0, 1]$ and the measurement action space $\mathcal{A}$ is quantized into a size-$Q_{a}$ set to make the minimization over $\mathcal{A}$ tractable.
To store expected future error functions $ \clbrsv{ R_S }{ S \subseteq [N] } $ and a set of best measurement action functions $ \clbrsv{ A_S }{ S \subseteq [N] } $, the memory complexity is $\mathrm{O}(2^N Q_p)$.
Besides, each value is obtained from a minimization over $S \times \mathcal{A}$, so the total computation is of complexity $\mathrm{O}(2^N Q_p N Q_a)$.
The number of DP functions $A_S$ and $R_S$ is $2^N$ because the order of measurement matters in general. 
However, DP still represents a speedup over the case of naive exhaustion for all possible orders, which has complexity $N!$. 

If all the qubits are identical copies, then the ordering is immaterial, and the different subsets $S$ with same size correspond to the same case.
Therefore, in this scenario the memory complexity is $\mathrm{O}(N Q_p)$ and the computation complexity is $\mathrm{O}(N Q_p Q_a)$.


Since there are only two possible states $ \rhohat_{+} $ and $ \rhohat_{-} $, we can also use log-likelihood ratios (LLR) to describe the probabilities.
This parameterization simplifies the computation of CSPs.
For $j \in [N]$, define
\begin{IEEEeqnarray}{rCl}
    \ell_0^\sigma(q) & \triangleq & \ln \left( \frac{q}{1 - q} \right), \\
    \ell_j^\sigma(q, \mathbf{a}_{[j]}^\sigma, \mathbf{d}_{[j]}^\sigma) & \triangleq & \ln \left( \frac{ C_j^\sigma(q, \mathbf{a}_{[j]}^\sigma, \mathbf{d}_{[j]}^\sigma) }{ 1 - C_j^\sigma(q, \mathbf{a}_{[j]}^\sigma, \mathbf{d}_{[j]}^\sigma) } \right), \\
    \tilde{\ell}_{\sigma(j)}(\mathbf{a}_{\sigma(j)}, d_{\sigma(j)}) & \triangleq & \ln \left( \frac{ \mathbb{P} \rdbrsv{ d_{\sigma(j)} }{ \rhohat_{+}, \mathbf{a}_{\sigma(j)}} }{ \mathbb{P} \rdbrsv{ d_{\sigma(j)} }{ \rhohat_{-}, \mathbf{a}_{\sigma(j)}} } \right).
\end{IEEEeqnarray}
It is easy to check that
\begin{IEEEeqnarray*}{rCl}
    \exp \rdbrs{ \ell_j^\sigma \rdbrs{ q, \mathbf{a}_{[j]}^\sigma, \mathbf{d}_{[j]}^\sigma } } & = & \frac{ C_j^\sigma \rdbrs{ q, \mathbf{a}_{[j]}^\sigma, \mathbf{d}_{[j]}^\sigma } }{ 1 - C_j^\sigma \rdbrs{ q, \mathbf{a}_{[j]}^\sigma, \mathbf{d}_{[j]}^\sigma } } \\
    &=& \frac{ \mathbb{P} \rdbrsv{ d_{\sigma(j)} }{ \rhohat_{+}, \mathbf{a}_{\sigma(j)}} \, C_{j-1}^\sigma \rdbrs{ q, \mathbf{a}_{[j-1]}^\sigma, \mathbf{d}_{[j-1]}^\sigma } }{ \mathbb{P} \rdbrsv{ d_{\sigma(j)} }{ \rhohat_{-} , \mathbf{a}_{\sigma(j)}} \, \rdbrs{ 1 - C_{j-1}^\sigma \rdbrs{ q, \mathbf{a}_{[j-1]}^\sigma, \mathbf{d}_{[j-1]}^\sigma } } } \\
    &=& \exp \rdbrs{ \tilde{\ell}_{\sigma(j)}(\mathbf{a}_{\sigma(j)}, d_{\sigma(j)}) } \cdot \exp \rdbrs{ \ell_{j-1}^\sigma \rdbrs{ q, \mathbf{a}_{[j-1]}^\sigma, \mathbf{d}_{[j-1]}^\sigma } }.
\end{IEEEeqnarray*}
This yields the simplified recursive equation
\begin{equation}
\label{top_down_recursion_llr_order}
    \ell_j^\sigma \rdbrs{ q, \mathbf{a}_{[j]}^\sigma, d_{[j]}^\sigma } = \tilde{\ell}_{\sigma(j)}(\mathbf{a}_{\sigma(j)}, d_{\sigma(j)}) + \ell_{j-1}^\sigma \rdbrs{ q, \mathbf{a}_{[j-1]}^\sigma, \mathbf{d}_{[j-1]}^\sigma }. 
\end{equation}

\subsection{Results for Qubits and Qutrits}

In the most general case, some properties of the algorithm remain unknown. 
For the special case where $\hat{\rho}_{\pm}$ are both pure, Theorem~\ref{thm:pure_states} shows that the optimal adaptive strategy consists of binary projective measurements and that its performance is unaffected by subsystem ordering. 
Therefore, a natural question to ask is whether adaptive binary projective measurements are always sufficient for general states. 
Additionally, we address the question of whether subsystem ordering affects the probability of success when optimization is done over all ``reasonable'' adaptive protocols.

\subsection*{Qubit Results}

We address the question of ordering by first demonstrating analytically that ordering can make a difference for a specific subset of candidate states when $N=2$. Then we perform experiments to show a small but nontrivial difference for more general tensor product states. 

First, we consider candidate states of the form 

\begin{align*}
    \rho_{+} &= \begin{pmatrix}
    1-x & 0 \\
    0 & x
    \end{pmatrix} \otimes \ket{\theta}\bra{\theta}, \\
    \rho_{-} &= \begin{pmatrix}
    x & 0 \\
    0 & 1-x
    \end{pmatrix} \otimes \ket{-\theta}\bra{\theta}.
\end{align*}

Measuring the subsystems in the best order (diagonal matrices first followed by $\ket{\pm \theta} \bra{\pm \theta}$) is equivalent to updating the prior from $q = \frac{1}{2}$ to $x$ and then implementing a Helstrom measurement on the second subsystem with the updated prior. The resulting probability of success is optimal and performs as well as a composite Helstrom measurement on both subsystems, namely, $P_{\mathrm{succ, best}} = \frac{1}{2}(1 + \sqrt{1 - 4(1-x)x\cos^{2}(2 \theta)})$. 

Measuring in the reverse order, the probability of success is $P_{\mathrm{succ, worst}} = \text{max} \{x, 1-x, \frac{1}{2}(1 + \sqrt{1 - \frac{1}{2} \cos^{2}(2 \theta)}) \}$ given that the diagonal subsystems are always optimally measured in the computational basis regardless of previous information. Thus, there is in general a difference between the best and worst ordering. 

We then perform experiments when $\hat{\rho}_{\pm}^{(j)}$ are all real qubit states. The set of measurements $\mathcal{A}_{\mathrm{qubit}}$ is taken to be the standard action space of real orthogonal projectors~\cite{Higgins-physreva11}  
\begin{align}
\mathcal{A}_{\mathrm{qubit}} \triangleq \left\{ \{ \ket{\phi}\bra{\phi}, \ket{\phi^{\perp}} \bra{\phi^{\perp}} \} \colon  \phi \in \left[ 0, \frac{\pi}{2} \right] \right\}
\end{align}
where we quantize $\phi$ into $Q_{\phi} = 128$ equally spaced points. 
The experimental setup is as follows:

\begin{enumerate}
\item   Choose a set of depolarizing parameters $\mathcal{S}_{\mathrm{dep}} = \{0, 0.05, 0.1, 0.2, 0.3, 0.4, 0.5, 0.6, 0.7, 0.8, 0.9, 1\}$ and number of trials $n_{\mathrm{trial}} = 1000$.
\item   Generate $\theta_\pm^{(t, j)} \in (0, 2 \pi)$ uniformly, where $t \in [n_{\mathrm{trial}}]$ denotes the trial index, and $j = 1, 2, \ldots, 7$ denotes the subsystem index.
\item   For each $\gamma \in \mathcal{S}_{\mathrm{dep}}$ and $N \in \{3, 4, 5, 6, 7\}$, define the corresponding qubit quantum states: 
        \begin{align}
            \rhohat_{\pm}(\gamma, t, N) \triangleq \bigotimes_{j=1}^{N} \left( (1-\gamma) \ketbra{\theta_\pm^{(t, j)}} + \frac{\gamma}{2}I \right).
        \end{align}
\item   For each $\hat{\rho}_{\pm}(\gamma, t, N)$ perform two separate optimizations corresponding to the best and worst ordering respectively, where the corresponding future risk functions are
        \begin{align}
            R_{S, \text{best}} \Big( p, \{\rho_{\pm}(\gamma, t, N)\} \Big) & \triangleq \min_{(k, \mathbf{a}_{k}) \in S \times \mathcal{A}} \sum_{d_{k} \in \{+, -\}} \mathcal{L}(p, \mathbf{a}_{k}, d_{k}) \cdot R_{S \backslash \{k\}}(\mathcal{P}(p, \mathbf{a}_{k}, d_{k})), \\
            R_{S, \text{worst}} \Big( p, \{\rho_{\pm}(\gamma, t, N) \} \Big) & \triangleq  \max_{k \in S}  \min_{\mathbf{a}_{k} \in \mathcal{A}}  \sum_{d_{k} \in \{+, -\}}  \mathcal{L}(p, \mathbf{a}_{k}, d_{k}) \cdot R_{S \backslash \{k\}}(\mathcal{P}(p, \mathbf{a}_{k}, d_{k})).
        \end{align}
\item   For $\gamma \in \mathcal{S}_{\mathrm{dep}}$ and $N \in \{3, 4, 5, 6, 7\}$, given an order of ``best'' or ``worst'', denote:
        \begin{align}
            P_{\mathrm{succ,order}}(N, \gamma) \triangleq \frac{1}{n_{\mathrm{trial}}} \sum_{t=1}^{n_{\mathrm{trial}}} P_{\mathrm{s,order}} \left( \rhohat_{\pm}(\gamma, t, N) \right)
        \end{align}
        where $P_{\mathrm{s,order}}(\rhohat_{\pm})$ indicates that we perform the MOODY algorithm with the specified ordering on states $\rhohat_{\pm}$.
\end{enumerate}

We plot $P_{\mathrm{succ,order}}(N=3, \gamma)$ as a function of $\gamma$ in Fig.~\ref{fig:N3qubit} and we also compare the difference $P_{\mathrm{succ,diff}}(N, \gamma) \triangleq P_{\mathrm{succ,best}}(N, \gamma) - P_{\mathrm{succ,worst}}(N, \gamma)$ for $N \in \{3,4,5,6,7\}$ in Fig.~\ref{fig:N3N4qubit}. 
From these results, we observe that the difference in probability of success with respect to ordering is not too large but it persists even when using the MOODY algorithm.

\begin{figure}
\begin{minipage}{.45\textwidth}
 \centering
  \scalebox{0.8}{
\begin{tikzpicture}

\begin{axis}[
legend cell align={left},
legend entries={{max},{min}},
legend style={at={(0.03,0.03)}, anchor=south west, draw=white!80.0!black},
tick align=outside,
tick pos=left,
x grid style={white!69.01960784313725!black},
xlabel={$\gamma$},
xmin=0, xmax=1.0,
y grid style={white!69.01960784313725!black},
ylabel={$P_{\text{succ,order}}(N=3, \gamma)$},
ymin=0.5, ymax=1.0,
xmajorgrids,
ymajorgrids
]
\addplot [semithick, red, mark=x, mark options={solid}]
table [row sep=\\]{%
0	0.962347800516396 \\
0.05	0.940613006009432 \\
0.1	0.920550345039296 \\
0.2	0.879251334478182 \\
0.3	0.836485303920521 \\
0.4	0.791947996988837 \\
0.5	0.74579084556548 \\
0.6	0.698453936000447 \\
0.7	0.64966395363632 \\
0.8	0.600336423495641 \\
0.9	0.550193575820532 \\
1	0.5 \\
};
\addplot [semithick, blue, mark=o, mark options={solid}]
table [row sep=\\]{%
0	0.962327874262567 \\
0.05	0.936890365038105 \\
0.1	0.916797880873348 \\
0.2	0.876118448665831 \\
0.3	0.833993543299578 \\
0.4	0.790136523310552 \\
0.5	0.74461831427777 \\
0.6	0.69778529849302 \\
0.7	0.64936860177256 \\
0.8	0.600242642184328 \\
0.9	0.550180886841754 \\
1	0.5 \\
};
\end{axis}

\end{tikzpicture}
  }
  \caption{\label{fig:N3qubit}Comparison of probabilities $P_{\mathrm{succ,best}}(N=3, \gamma)$ and $P_{\mathrm{succ,worst}}(N=3, \gamma)$ as a function of the depolarising parameter $\gamma$ over $1000$ trials. Although $P_{\mathrm{succ,best}}(N=3, \gamma) \ne P_{\mathrm{succ,worst}}(N=3, \gamma)$, the relative difference is small. 
}
\end{minipage}
\begin{minipage}{.05\textwidth}

\hspace*{0.1cm}

\end{minipage}
\begin{minipage}{.45\textwidth}
 \centering
  \scalebox{0.8}{
\begin{tikzpicture}

\definecolor{color0}{rgb}{0.75,0.75,0}

\begin{axis}[
legend cell align={left},
legend entries={{$N = 3$},{$N = 4$},{$N = 5$},{$N = 6$},{$N = 7$}},
legend style={draw=white!80.0!black},
tick align=outside,
tick pos=left,
x grid style={white!69.01960784313725!black},
xlabel={$\gamma$},
xmin=0, xmax=1,
y grid style={white!69.01960784313725!black},
y label style={at={(axis description cs:-0.1,.5)},rotate=0},
ylabel={$P_{\mathrm{succ,diff}}(N, \gamma)$},
ymin=-0, ymax=0.004,
scaled ticks=false, 
yticklabel style={/pgf/number format/fixed, /pgf/number format/fixed zerofill, /pgf/number format/precision=3},
ylabel near ticks,
xmajorgrids,
ymajorgrids
]
\addplot [semithick, red, mark=+, mark options={solid}]
table [row sep=\\]{%
0	1.99262538294054e-05 \\
0.05	0.00372264097132691 \\
0.1	0.0037524641659481 \\
0.2	0.00313288581235127 \\
0.3	0.00249176062094281 \\
0.4	0.00181147367828527 \\
0.5	0.00117253128770989 \\
0.6	0.000668637507426517 \\
0.7	0.00029535186375973 \\
0.8	9.37813113134345e-05 \\
0.9	1.26889787780415e-05 \\
1	0 \\
};
\addplot [semithick, blue, mark=o, mark options={solid}]
table [row sep=\\]{%
0	2.08230154389311e-05 \\
0.05	0.00285503764683881 \\
0.1	0.00301350133844469 \\
0.2	0.00310997547470671 \\
0.3	0.00279520660137766 \\
0.4	0.00222801408010487 \\
0.5	0.00154919423506417 \\
0.6	0.000911906052126876 \\
0.7	0.000438089652745832 \\
0.8	0.000167058459972891 \\
0.9	2.8985247537805e-05 \\
1	0 \\
};
\addplot [semithick, black, mark=x, mark options={solid}]
table [row sep=\\]{%
0	2.02119260206413e-05 \\
0.05	0.00230827284031732 \\
0.1	0.00256442926009415 \\
0.2	0.0027024660737468 \\
0.3	0.00252611545582526 \\
0.4	0.00207577080157906 \\
0.5	0.00147808150162609 \\
0.6	0.000893375094858095 \\
0.7	0.000437789934260246 \\
0.8	0.000183194564752354 \\
0.9	3.58570956070503e-05 \\
1	0 \\
};
\addplot [semithick, green!50.0!black, mark=triangle, mark options={solid}]
table [row sep=\\]{%
0	1.62275982105164e-05 \\
0.05	0.00180564930927718 \\
0.1	0.00214488617176989 \\
0.2	0.00240403643818932 \\
0.3	0.00240141152814111 \\
0.4	0.00209246598010526 \\
0.5	0.00156622847520571 \\
0.6	0.0009496832798358 \\
0.7	0.000502637250013094 \\
0.8	0.000214697735505087 \\
0.9	4.98326967000029e-05 \\
1	0 \\
};
\addplot [semithick, magenta, mark=square, mark options={solid}]
table [row sep=\\]{%
0	1.12520977285646e-05 \\
0.05	0.00136051877900389 \\
0.1	0.00175752805203422 \\
0.2	0.00208210363994343 \\
0.3	0.00213158129828006 \\
0.4	0.00192451789558135 \\
0.5	0.00148992603793296 \\
0.6	0.00098609013051254 \\
0.7	0.000511346708193527 \\
0.8	0.000236473279000671 \\
0.9	6.10318146242461e-05 \\
1	0 \\
};
\end{axis}

\end{tikzpicture}
  }
  \caption{\label{fig:N3N4qubit}Comparison of difference in maximum and minimum probability of success, $P_{\mathrm{succ,diff}}(N, \gamma)$, as a function of the depolarizing parameter $\gamma$ over $1000$ trials for $N=3, 4, 5, 6, 7$. 
}
\end{minipage}
\end{figure}

\subsection*{Qutrit Results}

Finally, we investigate whether restricting the action space to binary projectors is sufficient for non-qubit states, in particular for qutrit states.

\begin{definition}
Action space $\mathcal{A}$ is {\emph sufficient} for state space $\mathcal{H}$ if and only if for all $\hat{\rho}_{\pm} \in \mathcal{H}$ and $q \in [0, 1]$, 
\begin{align*}
    P_{\mathrm{succ}, \mathcal{A}}(q, \rho_{\pm}) = P_{\mathrm{succ}, \mathcal{A}_{\mathrm{all}}}(q, \rho_{\pm}), 
\end{align*}
where $\mathcal{A}_{\mathrm{all}}$ is the set of all quantum measurements of appropriate dimension, i.e., $\text{dim}(\rhohat_{\pm})$ and $P_{\mathrm{succ}, \mathcal{A}}(q, \rho_{\pm})$ is the probability of success of the order-optimized MOODY algorithm for a given action space $\mathcal{A}$. 
\end{definition}

For pure states, Theorem~\ref{thm:pure_states} confirms that binary projectors are sufficient, and by the definition of the Helstrom measurement, binary projectors are additionally sufficient whenever $N = 1$.

We show that binary projective measurements are not sufficient for general state spaces. 
To this aim, we define $\mathcal{H}_{\mathrm{qutrit}}$ to be the space of depolarized, real qutrit states and define the action space of real binary (ternary) measurements as $\mathcal{A}_{\mathrm{b}}$ ($ \mathcal{A}_{\mathrm{t}}$).
   \begin{align}
         \mathcal{A}_{\mathrm{b}} & \triangleq \Big\{  \{ \Pi_{j}^{\mathrm{b}} \}_{j=1}^{2} \ \Big| \ \Pi_{j}^{\mathrm{b}} \Pi_{j'}^{\mathrm{b}}= \delta_{j, j'} \Pi_{j}^{\mathrm{b}} \  \forall \ j, j' \in \{1, 2\}, \ \text{rank}(\Pi_{1}^{\mathrm{b}}) =2, \ \text{rank}(\Pi_{2}^{\mathrm{b}}) = 1 \ \Big\}, \\
        \mathcal{A}_{\mathrm{t}} & \triangleq \Big\{  \{ \Pi_{j}^{\mathrm{t}} \}_{j=1}^{3} \ \Big| \ \Pi_{j}^{\mathrm{t}} \Pi_{j'}^{\mathrm{t}}= \delta_{j, j'} \Pi_{j}^{\mathrm{t}}, \ \text{rank}(\Pi_{j}^{\mathrm{t}}) = 1 \ \forall \ j, j' \in \{1, 2, 3\} \Big\}.
    \end{align} 
Note that it is sufficient to consider real quantum measurements, as $\text{Tr}[\rho\, \Pi] = \text{Tr}[\rho \, \text{Re}(\Pi)]$ for any Hermitian projector $\Pi$ and so the resulting statistics will be invariant upon taking only the real part of any measurement set. 
Subject to the constraint that each subsystem may be measured only once, any ternary set of orthogonal projectors may be chosen to have all elements rank $1$ because any rank $2$ or $3$ element can be viewed as grouping the corresponding rank $1$ projectors post-measurement.

Then, parameterizing the action spaces is equivalent to generating (with some quantization) all orthonormal bases $\{ \ket{u_{1}}, \ket{u_{2}}, \ket{u_{3}} \}$ and, for each basis, defining the corresponding ternary POVM $ \Pi^{\mathrm{t}}\Big(\{ \ket{u_{1}}, \ket{u_{2}}, \ket{u_{3}} \} \Big)$ and three corresponding binary POVMs  $\Pi^{\mathrm{b}, k}\Big(\{ \ket{u_{1}}, \ket{u_{2}}, \ket{u_{3}} \} \Big)$, for $k \in \{1, 2, 3\}$ as follows:
\begin{align}
    \Pi^{\mathrm{t}}\Big(\{ \ket{u_{1}}, \ket{u_{2}}, \ket{u_{3}} \} \Big) & \triangleq \{ \ket{u_{1}}\bra{u_{1}}, \ \ket{u_{2}}\bra{u_{2}}, \ \ket{u_{3}}\bra{u_{3}} \}, \\
    \Pi^{\mathrm{b}, k}\Big(\{ \ket{u_{1}}, \ket{u_{2}}, \ket{u_{3}} \} \Big) & \triangleq \{ \sum_{l \ne k} \ket{u_l}\bra{u_l}, \ \ket{u_k}\bra{u_k} \}, 
\end{align} 
We implement this quantization using the following steps:
\begin{enumerate}

\item We quantize the unit sphere by subdividing an icosahedron for $T$ steps according to vector $\vec{r}=[r_{1}, ..., r_{T}]$ such that at the $j^{th}$  step we subdivide each segment by $r_{j}$. Then according to the Euler characteristic of convex polyhedrons, the number of vertices after all subdivisions are complete is given by $10 \prod_{i=1}^T r_i^2 + 2 $, and we denote this set of vertices as $\text{Sub}(\vec{r})$.  


\item For each point $(x,y,z)$ in $\text{Sub}([2,2,2])$, convert to polar coordinates $(\phi, \theta)$ according to
\begin{align*}
x = \sin(\theta) \cos(\phi), \ \ y = \sin(\theta)\sin(\phi), \ \ z = \cos(\theta).
\end{align*}

\item For each pair $(\phi, \theta)$, define the rotation matrix $R(\phi, \theta)$ as
\begin{align*}
R(\phi, \theta) & \triangleq 
\begin{bmatrix}
-\sin(\phi) & \cos(\phi)\cos(\theta) & \cos(\phi)\sin(\theta) \\
\cos(\phi) & \sin(\phi)\cos(\theta) & \sin(\phi) \sin(\theta) \\
0 & - \sin(\theta) & \cos(\theta)
\end{bmatrix}.
\end{align*}

\item Choose $Q$ as the resolution on the equatorial plane, let $\omega \in \{ \frac{\pi q}{2Q} \}_{q=0}^{Q-1} $ and define
\begin{equation*}
    \vec{u}_1(\phi, \theta, \omega) 
    = R(\phi, \theta)
    \begin{bmatrix}
        \cos(\omega) \\ \sin(\omega) \\ 0
    \end{bmatrix}, \quad
    \vec{u}_2(\phi, \theta, \omega) 
    = R(\phi, \theta)
    \begin{bmatrix}
        -\sin(\omega) \\ \cos(\omega) \\ 0
    \end{bmatrix}, \quad
    \vec{u}_3(\phi, \theta, \omega)
    = R(\phi, \theta)
    \begin{bmatrix}
        0 \\ 0 \\ 1
    \end{bmatrix}.
\end{equation*}

\end{enumerate}

The implied action spaces $\mathcal{A}_{\mathrm{t}}$ and $\mathcal{A}_{\mathrm{b}}$ are then used to compare the probability of successful discrimination in the ternary and binary cases, respectively. 
Using the following procedure, we demonstrate that for general real depolarized qutrit states $\{ \hat{\rho}_{+}, \hat{\rho}_{-} \}$, $P_{\mathrm{succ}, \mathcal{A}_{\mathrm{b}}}(\frac{1}{2}, \hat{\rho}_{\pm}) < P_{\mathrm{succ}, \mathcal{A}_{\mathrm{t}}}(\frac{1}{2}, \hat{\rho}_{\pm})$, and hence binary projective measurements are not sufficient.

\begin{enumerate}
\item   Fix $N = 3$, and choose a set of allowed depolarizing parameters $\mathcal{S}_{\mathrm{dep}} = \{0.1, 0.2, 0.3, 0.4, 0.5, 0.6\}$ and number of trials $n_{\mathrm{trial}} = 1000$.
\item   Generate $\alpha_\pm^{(t, j)}, \beta_\pm^{(t, j)} \in (0, 1)$ uniformly, where $t \in [n_{\mathrm{trial}}]$ denotes the trial index, and $j = 1, 2, \ldots, N$ denotes the subsystem index.
        Set $\phi_\pm^{(t, j)} = 2\pi \alpha_\pm^{(t, j)}$ and $\theta=\arccos(1 - 2\beta_\pm^{(t, j)})$, such that
        \begin{align*}
            \ket{v(\phi, \theta)} = [\sin(\theta)\cos(\phi),\ \sin(\theta) \sin(\phi),\ \cos(\theta)] 
        \end{align*}
        is uniformly distributed over the set of all unit vectors.
\item   For each $\gamma \in \mathcal{S}_{\mathrm{dep}}$, define the corresponding qutrit quantum states 
        \begin{align}
            \rhohat_{\pm}(\gamma, t, N) \triangleq \bigotimes_{j=1}^{N} \left( (1-\gamma) \ketbra{v(\phi_\pm^{(t, j)}, \theta_\pm^{(t, j)})} + \frac{\gamma}{3}I \right).
        \end{align}
\item   For each $\hat{\rho}_{\pm}(\gamma, t, N)$ perform the MOODY algorithm for $\mathcal{A}_{\mathrm{b}}$ and $\mathcal{A}_{\mathrm{t}}$ for both best ordering and worst ordering.
\item   For $\gamma \in \mathcal{S}_{\mathrm{dep}}$, given an order of ``best'' or ``worst'', and given an action space $\mathcal{A} \in \{ \mathcal{A}_{\mathrm{b}}, \mathcal{A}_{\mathrm{t}} \}$, denote
        \begin{align}
            P_{\mathrm{succ,order}}(\gamma, \mathcal{A}) = \frac{1}{n_{\mathrm{trial}}} \sum_{t=1}^{n_{\mathrm{trial}}} P_{\mathrm{s,order}} \left( \rhohat_{\pm}(\gamma, t, N), \mathcal{A} \right)
        \end{align}
        where $P_{\mathrm{s,order}}(\rhohat_{\pm}, \mathcal{A})$ indicates that we perform the MOODY algorithm over action space $\mathcal{A}$ with corresponding ordering on states $\rhohat_{\pm}$.
\end{enumerate}

We plot the results for all four methods in Fig.~\ref{fig:qutritall}, and compare the difference of the remaining three methods to the ternary, best ordering method ($P_{\mathrm{diff,order}}(\gamma, \mathcal{A}) = P_{\mathrm{succ,best}}(\gamma, \mathcal{A}_{t}) - P_{\mathrm{succ,order}}(\gamma, \mathcal{A})$) in Fig.~\ref{fig:qutritalldiff}.
We observe that the best ternary ordering is better than best binary ordering, and ordering still affects performance even in the MOODY algorithm. From this, we conjecture that for any action space and any adaptive approach, the order of subsystem measurement will affect the success probability. 
It remains an open question whether it is sufficient to consider $d$ rank $1$ orthogonal projectors for a state space $\mathcal{H}_{d}$  containing $d$-dimensional real quantum states.

\begin{figure}
\begin{minipage}{.45\textwidth}
 \centering
  \scalebox{0.8}{
\begin{tikzpicture}

\begin{axis}[
legend cell align={left},
legend entries={{ternary, best},{ternary, worst},{binary, best},{binary, worst}},
legend style={at={(0.03,0.03)}, anchor=south west, draw=white!80.0!black},
tick align=outside,
tick pos=left,
x grid style={white!69.01960784313725!black},
xlabel={$\gamma$},
xmin=0.1, xmax=0.6,
y grid style={white!69.01960784313725!black},
ylabel={$P_{\mathrm{succ,order}}(\gamma, \mathcal{A})$},
ymin=0.7, ymax=1.0,
xmajorgrids,
ymajorgrids
]
\addplot [semithick, red, mark=+, mark options={solid}]
table [row sep=\\]{%
0.1	0.962872207728902 \\
0.2	0.931796184894731 \\
0.3	0.895222732331908 \\
0.4	0.852506095954075 \\
0.5	0.804984681829628 \\
0.6	0.751844001875129 \\
};
\addplot [semithick, blue, mark=o, mark options={solid}]
table [row sep=\\]{%
0.1	0.960420532067588 \\
0.2	0.929157347393669 \\
0.3	0.892682545525711 \\
0.4	0.850359862841028 \\
0.5	0.803407211975032 \\
0.6	0.750834570798576 \\
};
\addplot [semithick, black, mark=x, mark options={solid}]
table [row sep=\\]{%
0.1	0.961326645249102 \\
0.2	0.928313604415174 \\
0.3	0.889997247343971 \\
0.4	0.845897151784528 \\
0.5	0.797483527430958 \\
0.6	0.744105577481252 \\
};
\addplot [semithick, green!50.0!black, mark=triangle, mark options={solid}]
table [row sep=\\]{%
0.1	0.958409766803538 \\
0.2	0.925036555689116 \\
0.3	0.886574554923471 \\
0.4	0.842499135877204 \\
0.5	0.794347283088988 \\
0.6	0.741313414134081 \\
};
\end{axis}

\end{tikzpicture}
  }
  \caption{\label{fig:qutritall}The average probability of success for the best and worst ordering using both ternary and binary projective measurements for qutrit product states when $N=3$. Results are averaged over $1000$ trials. 
}
\end{minipage}
\begin{minipage}{.05\textwidth}

\hspace*{0.1cm}

\end{minipage}
\begin{minipage}{.45\textwidth}
 \centering
 \vspace*{-0.05cm}
  \scalebox{0.8}{
 \begin{tikzpicture}

\begin{axis}[
legend cell align={left},
legend entries={{ternary, worst},{binary, best},{binary, worst}},
legend style={at={(0.97,0.47)}, anchor=north east, draw=white!80.0!black},
tick align=outside,
tick pos=left,
x grid style={white!69.01960784313725!black},
xlabel={$\gamma$},
xmin=0.1, xmax=0.6,
y grid style={white!69.01960784313725!black},
ylabel={$P_{\mathrm{diff, order}}(\gamma, \mathcal{A})$},
ymin=0, ymax=0.012,
scaled ticks=false, 
yticklabel style={/pgf/number format/fixed, /pgf/number format/fixed zerofill, /pgf/number format/precision=3},
xmajorgrids,
ymajorgrids
]
\addplot [semithick, red, mark=+, mark options={solid}]
table [row sep=\\]{%
0.1	0.00245167566131366 \\
0.2	0.00263883750106275 \\
0.3	0.0025401868061965 \\
0.4	0.00214623311304762 \\
0.5	0.00157746985459695 \\
0.6	0.0010094310765526 \\
};
\addplot [semithick, blue, mark=o, mark options={solid}]
table [row sep=\\]{%
0.1	0.00154556247979964 \\
0.2	0.00348258047955785 \\
0.3	0.00522548498793662 \\
0.4	0.00660894416954694 \\
0.5	0.00750115439867038 \\
0.6	0.00773842439387629 \\
};
\addplot [semithick, black, mark=x, mark options={solid}]
table [row sep=\\]{%
0.1	0.00446244092536408 \\
0.2	0.00675962920561535 \\
0.3	0.00864817740843704 \\
0.4	0.0100069600768716 \\
0.5	0.0106373987406408 \\
0.6	0.010530587741048 \\
};
\end{axis}

\end{tikzpicture}
  }
  \caption{\label{fig:qutritalldiff}Difference in average success probability for the various methods, namely, $P_{\mathrm{diff,order}}(\gamma, \mathcal{A})$ as a function of $\gamma$ when $N=3$. Results are averaged over $1000$ trials.}
\end{minipage}
\end{figure}

\section{Conclusion}
\label{sec:conclusion}

In this work, we investigated simple locally greedy and modified locally greedy algorithms as well as more general dynamic programming-based algorithms for quantum state discrimination when the given states are tensor products of $N$ arbitrary qubit or qutrit states.
We analytically proved that when the individual subsystems are pure states the simple locally greedy algorithm achieves the optimal performance of the joint $N$-system Helstrom measurement.
For the scenario of each subsystem containing identical copies of arbitrary qubit states, we demonstrated the plateau in the probability of success attained by the locally greedy algorithm with increasing $N$.
The reason for this plateau was discussed and an explicit bound derived for the success probability as a function of the channel depolarizing parameter and the initial prior. Based on these results, we introduced a modified locally greedy algorithm with strictly better performance and optimal discrimination in the large $N$ limit.  

For the general MOODY algorithm, we show that ordering of subsystems continues to affect the performance when the individual subsystems have distinct states.
Finally, for qutrit states we showed that binary projective measurements are inadequate to achieve optimal performance.
In future work we will complement the insights provided here using empirical results with rigorous analytical arguments.


\appendices

\section{Proof of Theorem~\ref{thm:pure_states}}
\label{sec:pure_states_proof}

We prove the statement by induction, first considering the base case where $N = 2$ and additionally specifying $q = \frac{1}{2}$. We leave $q$ implicit in the following until stated otherwise. Then the probability of success can be written as
\begin{align}
 P_{\mathrm{s, lg}}\left( \frac{1}{2}, \rhohat_{\pm} \right) & =  \sum_{d_{1} \in \{+, - \}}  \mathbb{P}(\rhohat =  \rhohat_{+}) \mathbb{P}\left( d_{1} \middle| \rhohat_{+} \right) \mathbb{P}\left( d_{2} = + \middle|  \rhohat_{+}, d_{1} \right) +  \sum_{d_{1} \in \{+, -\} } \mathbb{P}(\rhohat = \rhohat_{-}) \mathbb{P}\left( d_{1} \middle| \rhohat_{-} \right) \mathbb{P}\left( d_{2} = - \middle| \rhohat_{-}, d_{1} \right) \nonumber \\
 & = \frac{1}{2} \sum_{d_{1} \in \{+, - \}} \left[ \mathbb{P}\left( d_{1} \middle| \rhohat_{+} \right) \mathbb{P}\left( d_{2} = + \middle| \rhohat_{+}, d_{1} \right) +  \mathbb{P}\left( d_{1} \middle|  \rhohat_{-} \right) \mathbb{P}\left( d_{2} = - \middle| \rhohat_{-}, d_{1} \right) \right] \nonumber \\
 & = \mathbb{P}\left( d_{1}=+ \middle| \rhohat_{+} \right) \mathbb{P}\left( d_{2} = + \middle| \rhohat_{+}, d_{1}=+ \right) +  \mathbb{P}\left( d_{1}=+ \middle|  \rhohat_{-} \right) \mathbb{P}\left( d_{2} = - \middle| \rhohat_{-}, d_{1}=+ \right),
\end{align}
where the final equality follows from symmetry of probability outcomes for pure states, i.e.,
\begin{align*}
\mathbb{P}\left( d_{j} = + \middle| \rhohat_{+}, d_{1} = \pm \right) &= \mathbb{P}\left( d_{j} = - \middle| \rhohat_{-}, d_{1} = \mp \right).
\end{align*}
From the definition of the Helstrom measurement, we observe that
\begin{align}
\mathbb{P}\left( d_{1} = \pm \middle| \rhohat_{\pm} \right) = \frac{1}{2} (1 + \sin(\theta_{1})), \ 
\mathbb{P}\left( d_{1} = \pm \middle| \rhohat_{\mp} \right) = \frac{1}{2} (1 - \sin(\theta_{1})).
\end{align}
Hence the updated prior is $p_{1}(d_{1}) = \frac{1}{2} (1 + d_{1} \sin(\theta_{1}))$. 
Then according to the locally greedy algorithm, 
\begin{align}
\mathbb{P}\left( d_{2} \middle| \rhohat_{\pm}, d_{1} \right) & = 
\begin{cases}
1- \trays{\Pi \left( p_{1}(d_{1}), j = 1 \right) \rhohat_{\pm}} & \ \text{if} \ d_{2} = +, \\
\trays{\Pi \left( p_{1}( d_{1}), j = 1 \right) \rhohat_{\pm}} & \ \text{if} \ d_{2} = -.
\end{cases}
\end{align}
Equations 2.13-2.14 from ~\cite{Helstrom-jsp69} provide a solution for $\mathbb{P}\left( d_{2} \middle| \rhohat_{\pm}, d_{1} \right)$. Upon simplifying, we observe:
\begin{align}
\mathbb{P}\left( d_{2} = \pm \middle| \rhohat_{+}, d_{1} = \pm \right) = \frac{1}{2}\left( 1 \pm \frac{\sin^{2}{\theta_{2}} \pm \cos^{2}{\theta_{2}} \sin{\theta_{1}}}{\sqrt{\cos^{2}(\theta_{2}) \sin^{2}(\theta_{1}) + \sin^{2}(\theta_{2})}} \right).
\end{align}
Again using the symmetry property we have
\begin{align}
\mathbb{P}\left( d_{2} = \pm \middle| \rhohat_{-}, d_{1} = \pm \right) = \frac{1}{2}\left( 1 \mp \frac{\sin^{2}{\theta_{2}} \pm \cos^{2}{\theta_{2}} \sin{\theta_{1}}}{\sqrt{\cos^{2}(\theta_{2}) \sin^{2}(\theta_{1}) + \sin^{2}(\theta_{2})}} \right).
\end{align}
Upon substitution we obtain
\begin{align}
P_{\mathrm{s, lg}}\left( \frac{1}{2}, \rhohat_{\pm} \right) & = \mathbb{P}\left( d_{1} = + \middle| \rhohat_{+} \right) \mathbb{P}\left( d_{2} = + \middle| \rhohat_{+}, d_{1}=+ \right) + \mathbb{P}\left( d_{1} = + \middle|  \rhohat_{-} \right) \mathbb{P}\left( d_{2} = - \middle| \rhohat_{-}, d_{1} = + \right) \nonumber \\
  & = \frac{1}{2}\left( 1 + \sqrt{\cos^{2}(\theta_{2}) \sin^{2}(\theta_{1}) + \sin^{2}(\theta_{2})} \right)  \nonumber \\
  & = \frac{1}{2}\left( 1 + \sqrt{1 - \cos^{2}(\theta_{1}) \cos^{2}(\theta_{2})} \right).
\end{align}
For the inductive step, we define a new variable $\tilde{\theta} \in [0, \pi]$ such that $\cos^{2}(\tilde{\theta}) \triangleq \Pi_{i=1}^{N-1} \cos^{2}(\theta_{i})$. 
Then by assumption
\begin{align}
P_{\mathrm{s, lg}}\left( \frac{1}{2}, \rhohat_{\pm}^{(1, \ldots, N-1)} \right) & = \frac{1}{2} \left( 1 + \sqrt{1 - \Pi_{i=1}^{N-1} \cos^{2}(\theta_{i})} \, \right) = \frac{1}{2} \left( 1 + \sqrt{1 - \cos^{2}(\tilde{\theta})} \, \right).
\end{align}
We then apply the previously shown statement for $N = 2$, letting the first subsystem now be the combined subsystems $1, 2, \ldots, N-1$, i.e., $\rhohat_{\pm}^{(1, \ldots, N-1)}$.
\begin{align}
P_{\mathrm{s, lg}}\left( \frac{1}{2}, \rhohat_{\pm} \right) & = \frac{1}{2} \left( 1 + \sqrt{1 -  \cos^{2}(\tilde{\theta}) \cos^{2}(\theta_{N})} \, \right) \nonumber \\
  & = \frac{1}{2} \left( 1 + \sqrt{1 - \Pi_{i=1}^{N-1} \cos^{2}(\theta_{i}) \cos^{2}(\theta_{N})} \, \right) \\
  & = \frac{1}{2} \left( 1 + \sqrt{1 - \Pi_{i=1}^{N} \cos^{2}(\theta_{i})} \, \right).
\end{align}

Now we consider the case of general priors. 
We can artificially rearrange this problem so that it is mathematically equivalent to a new quantum state discrimination problem between two transformed states $\hat{\rho'}_{\pm}$. 
We start by defining $\theta_{0}$ such that $q = \frac{1}{2}(1 + \sin(\theta_{0}))$. 
For pure states, we have $P_{\mathrm{s, h}}(q, \rhohat_{\pm}) = P_{\mathrm{s, h}}(1-q, \rhohat_{\pm})$, so
\begin{align*}
P_{\mathrm{s, h}}(q, \rhohat_{\pm}) & = \frac{1}{2} \left[ P_{\mathrm{s, h}} \left( \frac{1 + \sin\theta_{0}}{2}, \rhohat_{\pm} \right) + 
P_{\mathrm{s, h}} \left( \frac{1 - \sin\theta_{0}}{2}, \rhohat_{\pm} \right) \right] \\
 & = P_{\mathrm{s, lg}}\left( \frac{1}{2}, \rhohat_{\pm}^{(0)} \otimes \rhohat_{\pm} \right) \\
 & = \frac{1}{2} \Bigg[ \mathbb{P}\left( d_0 = + \middle| \rhohat^{(0)} = \rhohat_{+}^{(0)} \right) \cdot \mathbb{P}\left( d_N = + \middle| d_{0} = +, \rhohat = \rhohat_{+} \right) \\
 & \qquad +  \mathbb{P}\left( d_0 = + \middle| \rhohat^{(0)} = \rhohat_{-}^{(0)} \right) \cdot \mathbb{P}\left( d_N = - \middle| d_{0}= +, \rhohat = \rhohat_{-} \right) \\
 & \qquad +  \mathbb{P}\left( d_0 = - \middle| \rhohat^{(0)} = \rhohat_{+}^{(0)} \right) \cdot \mathbb{P}\left( d_N = + \middle| d_{0} = -, \rhohat = \rhohat_{+} \right) \\
 & \qquad + \mathbb{P}\left( d_0 = - \middle|  \rhohat^{(0)} = \rhohat_{-}^{(0)} \right) \cdot \mathbb{P}\left( d_N = - \middle| d_{0} = -, \rhohat = \rhohat_{+} \right) \Bigg], 
\end{align*}
where we define the newly appended states $\rhohat_{\pm}^{(0)} \triangleq \ketbra{\theta_{0, \pm}}$ such that $|\braket{\theta_{0, +}}{\theta_{0, -}}|^{2}= \cos^{2}(\theta_{0})$. 
Here $\mathbb{P}(d_{N} \mid d_{0}, \text{state})$ denotes the probability of obtaining $d_{N}$ as the measurement result on the $N^{\mathrm{th}}$ subsystem given the updated prior $P_{1}(q_{0} = \frac{1}{2}, d_{0})$ and state, with all local measurements determined by the locally greedy algorithm.

Since we have restricted all quantum subsystems to be in pure states, we can simplify through symmetry as follows.
\begin{align*}
\mathbb{P}\left( d_N = + \middle| d_{0} = -, \rhohat = \rhohat_{+} \right) & = \mathbb{P}\left( d_N = - \middle| d_{0} = +, \rhohat = \rhohat_{-} \right)  \\
\mathbb{P}\left( d_N = - \middle| d_{0} = -, \rhohat = \rhohat_{-} \right) & = \mathbb{P}\left( d_N = + \middle| d_{0} = +, \rhohat = \rhohat_{+} \right).
\end{align*}
Substituting these properties we obtain
\begin{align*}
P_{\mathrm{s, h}}(q, \rhohat_{\pm}) & =  \mathbb{P}\left( d_0 = + \middle| \rhohat^{(0)} = \rhohat_{+}^{(0)} \right) \cdot \mathbb{P}\left( d_N = + \middle| d_{0} = +, \rhohat = \rhohat_{+} \right) \\
  & \qquad  +  \mathbb{P}\left( d_0 = + \middle| \rhohat^{(0)} = \rhohat_{-}^{(0)} \right) \cdot \mathbb{P}\left( d_N = - \middle| d_{0} = +, \rhohat = \rhohat_{-} \right) \\
&= q \cdot  \mathbb{P}\left( d_N = + \middle| d_{0} = +, \rhohat = \rhohat_{+} \right) + (1-q) \cdot \mathbb{P}\left( d_N = - \middle| d_{0} = +, \rhohat = \rhohat_{-} \right) \\
  & = P_{\mathrm{s, lg}}(q, \rhohat_{\pm}).
\end{align*}
The last equality follows from noting that the updated prior $P_{1}(q_{0} = \frac{1}{2}, +) = \frac{1}{2}(1 + \sin(\theta_{0})) = q$. Thus, the probability of success is equivalent under both the joint $N$-system Helstrom measurement and the locally greedy method for pure states.  \hfill \IEEEQEDhere

\section{Further Properties of the Locally Greedy Method }
\label{sec:additional_lgm}

We show the plateau remains in the order-optimized locally greedy algorithm by generalizing the experiment from identical copies to the case where the subsystems are distinct.
The primary change is that we now sample states parameterized by $\theta_\pm^{(t, j)}$ so that each subsystem in both $\rhohat_{+}$ and $\rhohat_{-}$ can have (potentially) distinct copies.
Also, the vector of success probabilities is altered accordingly.
\begin{enumerate}

\item Choose a set of depolarizing parameters and number of trials.
Again we set $\mathcal{S}_{\mathrm{dep}} = \{0.01, 0.05, 0.1, 0.3\}$ and $n_{\mathrm{trial}} = 1000$.

\item Generate $\theta_\pm^{(t, j)} \in (0, 2 \pi)$ uniformly, where $t \in [n_{\mathrm{trial}}]$ denotes the trial index, and $j = 1, 2, \ldots, 12$ denotes the subsystem index.

\item For each $\gamma \in \mathcal{S}_{\mathrm{dep}}$ and $N = 1, 2, \ldots, 12$, define the corresponding qubit quantum states 
\begin{align}
\rhohat_{\pm}(\gamma, t, N) \triangleq \bigotimes_{j=1}^{N} \left( (1-\gamma) \ketbra{\theta_\pm^{(t, j)}} + \frac{\gamma}{2}I \right).
\end{align}

\item For all $\gamma \in \mathcal{S}_{\mathrm{dep}}$ and all $N = 1, 2, \ldots, 12$, denote
\begin{align}
    P_{\mathrm{succ}}(N, \gamma) = \frac{1}{n_{\mathrm{trial}}} \sum_{t=1}^{n_{\mathrm{trial}}} P_{\mathrm{s,lg}} \left( \rhohat_{\pm}(\gamma, t, N) \right),
\end{align}
where $P_{\mathrm{s,lg}}(\rhohat_{\pm})$ indicates that we perform the locally greedy algorithm on states $\rhohat_{\pm}$.





\end{enumerate}

We plot the results of this experiment in Fig.~\ref{fig:plateau}.
When the subsystems are not identical copies, we notice that the plateau is much higher when compared to the case of identical copies.
At first glance this appears to violate the bound obtained in Corollary~\ref{cor:succ_prob_bound}, since here the pair of states in each subsystem are pure states depolarized with the same parameter $\gamma$, as required in the hypothesis of Corollary~\ref{cor:succ_prob_bound}.
However, the reason for the higher plateau is as follows.
The algorithm orders the subsystems in such a way that the credulity can be updated to be as close to $1-\frac{\gamma}{2}$ as possible (where we assume the states $\rho_{\pm}$ may be relabeled at any step to ensure the credulity is always greater than $\frac{1}{2}$).
In the next round, it is still possible to obtain one more non-trivial measurement, after which either the updated credulity exceeds $1 - \frac{\gamma}{2}$ and all subsequent rounds are trivial, or the credulity is lowered below the threshold and another measurement is permitted until the updated credulity again exceeds $1 - \frac{\gamma}{2}$. This permitted ``jump'' in credulity due to the final measurement explains why the value appearing as the plateau in Fig. ~\ref{fig:plateau} can be larger than $1 - \frac{\gamma}{2}$. 


The best ``jump'' beyond $1-\frac{\gamma}{2}$ is obtained when the states in that subsystem are an orthogonal pair of pure states subjected to the depolarizing channel and a measurement result which increases the credulity is attained, as formalized in the following lemma. \par

\begin{lemma}
\label{lem:success_bound}
Suppose that we are given one of two quantum states $\rhohat_{+} = \bigotimes_{j=1}^{N} \Big( (1- \gamma)\ketbra{\theta_{j, +}} + \frac{\gamma}{2} \mathbb{I} \Big)$ and $\rhohat_{-} = \bigotimes_{j=1}^{N} \Big( (1- \gamma)\ketbra{\theta_{j, -}} + \frac{\gamma}{2} \mathbb{I} \Big)$ where $\gamma$ is a  fixed depolarizing parameter. Then, an upper bound on the probability of success using the locally greedy method is given by
\begin{align*}
    P_{\mathrm{s, lg}}\big( \frac{1}{2}, \rhohat_{\pm} \big) \leq P_{\text{bound}}(\gamma) \equiv \frac{(1 - \frac{\gamma}{2})^{2}}{(1 - \frac{\gamma}{2})^{2} + (\frac{\gamma}{2})^{2}}.
\end{align*}
\end{lemma}
\begin{IEEEproof}
  We show that the probability of success is upper bounded by the maximal attainable credulity after $N$ measurements, and proceed by induction. For details, see Appendix~\ref{sec:success_bound_proof}.
\end{IEEEproof}

To illustrate the predictive value of this bound, we list the observed numerical asymptotic values found when $N=12$ for non-identical subsystems ($P_{\mathrm{obs}}(\gamma)$) and the predicted upper bound for $\gamma = 0.1, 0.3, 0.4, 0.5$ respectively:
\begin{align*}
\bigg\{ \big(P_{\mathrm{obs}}(0.1) = 0.9943, P_{\mathrm{bound}}(0.1) = 0.9972 \big), \big(P_{\mathrm{obs}}(0.3) = 0.9549, P_{\mathrm{bound}}(0.3) = 0.9698 \big), \\ \big(P_{\mathrm{obs}}(0.4) = 0.9198, P_{\mathrm{bound}}(0.4) = 0.9412 \big), \big(P_{\mathrm{obs}}(0.5) = 0.8732, P_{\mathrm{bound}}(0.5) = 0.9000 \big) \bigg\}.
\end{align*}



Finally, we compare the two scenarios for the specific value of the depolarizing parameter $\gamma = 0.3$ in Fig.~\ref{fig:plateaucomp}.
This plot shows the non-trivial advantage obtained from subsystems being distinct rather than copies of each other, which is the case most considered in the literature.
For the special case of $\gamma = 0$, we have shown in Theorem~\ref{thm:pure_states} that the order of subsystems does not matter and that the simple locally greedy algorithm itself achieves the optimal performance obtained with the joint $N$-system Helstrom measurement.

\begin{figure}
\begin{minipage}{.45\textwidth}
 \centering
  \scalebox{0.8}{
\begin{tikzpicture}[every pin/.style={fill=white}]

\begin{axis}[
legend cell align={left},
legend style={at={(0.97,0.03)}, anchor=south east, draw=white!80.0!black},
tick align=outside,
tick pos=left,
x grid style={white!69.01960784313725!black},
xlabel={$N$},
xmin=1, xmax=12,
y grid style={white!69.01960784313725!black},
ylabel={$P_{\textrm{succ}}(N, \gamma)$},
ymin=0.7, ymax=1.02,
xmajorgrids,
ymajorgrids,
legend style={at={(0.02,0.02)}, anchor=south west}
]
\addplot [semithick, red, mark=+, mark options={solid}]
table [row sep=\\]{%
1	0.811107269179847 \\
2	0.909995671732371 \\
3	0.956742463421262 \\
4	0.9777540510538 \\
5	0.987126780376899 \\
6	0.993208545164197 \\
7	0.995813392636815 \\
8	0.997501137307117 \\
9	0.998594129908456 \\
10	0.999182750224289 \\
11	0.999484467782538 \\
12	0.999784261938053 \\
};
\addlegendentry{$\gamma = 0.01$}

\addplot [semithick, blue, mark=o, mark options={solid}]
table [row sep=\\]{%
1	0.798537278208313 \\
2	0.89235525243052 \\
3	0.939122720290063 \\
4	0.962768798642153 \\
5	0.974959521893912 \\
6	0.98378125325614 \\
7	0.988581212220827 \\
8	0.991873659660947 \\
9	0.994212216628297 \\
10	0.995664367031094 \\
11	0.996574581054864 \\
12	0.997351886698924 \\
};
\addlegendentry{$\gamma = 0.05$}

\addplot [semithick, black, mark=x, mark options={solid}]
table [row sep=\\]{%
1	0.782824789820195 \\
2	0.870458875920156 \\
3	0.917826446661341 \\
4	0.944647388066996 \\
5	0.959940513778371 \\
6	0.971645395338663 \\
7	0.978642731666403 \\
8	0.983635212656461 \\
9	0.987290920413494 \\
10	0.989701387264015 \\
11	0.991342688555298 \\
12	0.992819248286669 \\
};
\addlegendentry{$\gamma = 0.1$}

\addplot [
    semithick, 
    green!50.0!black, 
    mark=triangle, 
    mark options={solid}
]
table [row sep=\\]{%
1	0.719974835725761 \\
2	0.784842058966874 \\
3	0.832228903573413 \\
4	0.865239920847174 \\
5	0.887701136993284 \\
6	0.906232956723757 \\
7	0.919126125618693 \\
8	0.929412348971475 \\
9	0.937407888658741 \\
10	0.943317414278849 \\
11	0.949174887992815 \\
12	0.951386090757464 \\
};
\addlegendentry{$\gamma = 0.3$}

\end{axis}

\node[pin=0:{%
    \begin{tikzpicture}[baseline,trim axis left,trim axis right]
    \begin{scope}[scale=0.5, every node/.append style={transform shape}]
    \begin{axis}[
    ticklabel style = {font=\Large},
    tick align=inside,
    x grid style={white!69.01960784313725!black},
    xmin=7.9, xmax=12.1,
    y grid style={white!69.01960784313725!black},
    ymin=0.93, ymax=1.01,
    ytick={0.96, 0.97, 0.98, 0.99, 1},
    xmajorgrids,
    ymajorgrids
    ]
    \addplot [ultra thick, red, mark=+, mark options={solid}, mark size=5pt]
    table [row sep=\\]{%
    1	0.811107269179847 \\
    2	0.909995671732371 \\
    3	0.956742463421262 \\
    4	0.9777540510538 \\
    5	0.987126780376899 \\
    6	0.993208545164197 \\
    7	0.995813392636815 \\
    8	0.997501137307117 \\
    9	0.998594129908456 \\
    10	0.999182750224289 \\
    11	0.999484467782538 \\
    12	0.999784261938053 \\
    };

    \addplot [ultra thick, blue, mark=o, mark options={solid}, mark size=5pt]
    table [row sep=\\]{%
    1	0.798537278208313 \\
    2	0.89235525243052 \\
    3	0.939122720290063 \\
    4	0.962768798642153 \\
    5	0.974959521893912 \\
    6	0.98378125325614 \\
    7	0.988581212220827 \\
    8	0.991873659660947 \\
    9	0.994212216628297 \\
    10	0.995664367031094 \\
    11	0.996574581054864 \\
    12	0.997351886698924 \\
    };

    \addplot [ultra thick, black, mark=x, mark options={solid}, mark size=5pt]
    table [row sep=\\]{%
    1	0.782824789820195 \\
    2	0.870458875920156 \\
    3	0.917826446661341 \\
    4	0.944647388066996 \\
    5	0.959940513778371 \\
    6	0.971645395338663 \\
    7	0.978642731666403 \\
    8	0.983635212656461 \\
    9	0.987290920413494 \\
    10	0.989701387264015 \\
    11	0.991342688555298 \\
    12	0.992819248286669 \\
    };

    \addplot [ultra thick, green!50.0!black, mark=triangle, mark options={solid}, mark size=5pt]
    table [row sep=\\]{%
    1	0.719974835725761 \\
    2	0.784842058966874 \\
    3	0.832228903573413 \\
    4	0.865239920847174 \\
    5	0.887701136993284 \\
    6	0.906232956723757 \\
    7	0.919126125618693 \\
    8	0.929412348971475 \\
    9	0.937407888658741 \\
    10	0.943317414278849 \\
    11	0.949174887992815 \\
    12	0.951386090757464 \\
    };

    \end{axis}
    \end{scope}
    \end{tikzpicture}%
}] at (2.3, 1.8) {};

\end{tikzpicture}

  }
  \caption{\label{fig:plateau} Comparison of probability of success for varying $\gamma$ in the distinct subsystems scenario, as a function of the number of available systems, $N$. Results are average over $1000$ trials. 
  }
\end{minipage}
\begin{minipage}{.05\textwidth}

\hspace*{0.1cm}

\end{minipage}
\begin{minipage}{.45\textwidth}
 \centering
  \scalebox{0.8}{
\begin{tikzpicture}

\begin{axis}[
legend cell align={left},
legend style={at={(0.03,0.03)}, anchor=south west, draw=white!80.0!black},
tick align=outside,
tick pos=left,
x grid style={white!69.01960784313725!black},
xlabel={$N$},
xmin=0.5, xmax=12.5,
y grid style={white!69.01960784313725!black},
ylabel={$P_{\text{succ}}(N, \gamma = 0.3)$},
ymin=0.48, ymax=1.02,
xmajorgrids,
ymajorgrids
]
\addplot [semithick, blue, mark=x, mark options={solid}]
table [row sep=\\]{%
1	0.723505519075038 \\
2	0.74063678311536 \\
3	0.77970051373649 \\
4	0.787692435350777 \\
5	0.797293897078216 \\
6	0.797754428716 \\
7	0.801694096037334 \\
8	0.803102698213465 \\
9	0.8007408774783 \\
10	0.801703427216124 \\
11	0.801600354385908 \\
12	0.802214548357316 \\
};
\addlegendentry{Identical Copies}

\addplot [semithick, red, mark=o, mark options={solid}]
table [row sep=\\]{%
1	0.719974835725761 \\
2	0.784842058966874 \\
3	0.832228903573413 \\
4	0.865239920847174 \\
5	0.887701136993284 \\
6	0.906232956723757 \\
7	0.919126125618693 \\
8	0.929412348971475 \\
9	0.937407888658741 \\
10	0.943317414278849 \\
11	0.949174887992815 \\
12	0.951386090757464 \\
};
\addlegendentry{Distinct Subsystems}
\end{axis}

\end{tikzpicture}
  }
  \caption{\label{fig:plateaucomp} Comparison of probability of success as a function of the number of available systems, $N$, for depolarizing parameter $\gamma = 0.3$. 
Results are averaged over $1000$ trials.}
\end{minipage}
\end{figure}

\section{Proof of Lemma~\ref{lem:success_bound}}
\label{sec:success_bound_proof}
Given the sequence of measurement results $d_{[N]}$, the probability of success is $\max( p_{N}, 1- p_{N})$ where $p_{N} = C_{N}^{\sigma}(q, 
\mathbf{a}_{[N]}^{\sigma}, \mathbf{d}_{[N]}^{\sigma})$ (and where $\mathbf{a}_{\sigma(j)} = \Pi(j, p_{j})$.) We suppose that at any step we may swap the labels of the composite states to enforce $p_{j} \geq \frac{1}{2}$ \ $\forall \ j$. Then, the probability of success is upper bounded by the maximal attainable probability, namely $P_{s, lg}(\frac{1}{2}, \rhohat_{\pm}) \leq \max_{\mathbf{d}_{[N]}^{\sigma} \in \mathcal{D}^{N}} \big( C_{N}^{\sigma}(\frac{1}{2}, \mathbf{d}_{[N]}^{\sigma}) \big)$. \\
\\
We then show by induction that $p_{j} \leq \frac{(1 - \frac{\gamma}{2})^{2}}{(1 - \frac{\gamma}{2})^{2} + (\frac{\gamma}{2})^{2}}$ \ $\forall \ j \in \{0, 1, ..., N\}$. For the base case,  $j = 0$ and the statement is trivially true as $p_{0} = q= \frac{1}{2}$. For the inductive step, we assume that the statement holds for $j \in \{0, 1, ..., N-1\}$ and show that it then also holds for $j+1$. First, consider the case where the updated prior at the $j$-th step exceeds the critical value $1 - \frac{\gamma}{2} \leq p_{j}$. Then, all future measurements are trivial and $p_{k} = p_{j}$ for all $k \geq j$. Thus, the inductive hypothesis again holds. \\
\\
Next, consider the case where $p_{j} \in [\frac{1}{2}, 1- \frac{\gamma}{2}]$. For simplicity, we define $p_{j+1}(p_{j}, \mathbf{d}_{j+1}^{\sigma}) \triangleq C_{j+1}^{\sigma}(q, 
\mathbf{a}_{[j+1]}^{\sigma}, \mathbf{d}_{[j+1]}^{\sigma})$ with $a_{\sigma(j)} = \Pi(j, p_{j})$. If the Helstrom measurement $\Pi_{hel} \equiv \Pi(j+1, p_{j})$ is trivial, then $p_{j} = p_{j+1}$ and the inductive hypothesis holds. In the case where $\Pi_{hel}$ is nontrivial, the prior will increase when $d_{j+1} = +$ and the new maximal credulity $p_{j+1}^{*}$ is defined as follows: 
\begin{align*}
    p_{j+1}^{*} &= \max\Big(p_{j+1}(p_{j}, \ \mathbf{d}_{j+1}^{\sigma}= +),  p_{j+1}(p_{j}, \ \mathbf{d}_{j+1}^{\sigma}= -) \Big) \\
    &= p_{j+1}(p_{j}, \ \mathbf{d}_{j+1}^{\sigma}= +) \\
    &= \frac{p_{j} \text{Tr}[\big( \mathbb{I} - \Pi_{h} \big) \big((1- \gamma)\ketbra{\theta_{j, +}} + \frac{\gamma}{2} \mathbb{I})]}{\text{Tr}[\big( \mathbb{I} - \Pi_{h} \big) \big(p_{j} (1- \gamma)\ketbra{\theta_{j, +}} +(1-p_{j})(1-\gamma)\ketbra{\theta_{j, -}} + \frac{\gamma}{2} \mathbb{I})]} \\
    &= \frac{(1 - \gamma) p_{j} \text{Tr}[\big( \mathbb{I} - \Pi_{h} \big) \ketbra{\theta_{j, +}}] + \frac{p_{j} \gamma}{2}}{(1- \gamma)\text{Tr}[\big( \mathbb{I} - \Pi_{h} \big) \big(p_{j} \ketbra{\theta_{j, +}} +(1-p_{j})\ketbra{\theta_{j, -}}] + \frac{\gamma}{2}} \\
    &= \frac{(1- \gamma)p_{j} x_{+} + \frac{p_{j} \gamma}{2}}{(1- \gamma)(p_{j} x_{+} + (1 - p_{j})x_{-}) + \frac{\gamma}{2}} \\
\end{align*}
for $x_{\pm} \equiv \text{Tr}[(\mathbb{I} - \Pi_{h}) \ketbra{\theta_{j, \pm}}] \in [0, 1]$. The third line follows from substituting into Bayes' law and simplifying. In the following, we derive an upper bound on $p_{j+1}^{*}$ and thus the success probability by optimizing over $x_{+}, x_{-}, p_{j}$ without placing any restrictions on whether the optimal set $\{x_{+}^{*}, x_{-}^{*}, p_{j}^{*}\}$ is actually physically realizable.

\begin{align*}
    p_{j+1}^{*} & \leq  \max_{x_{\pm} \in [0, 1]} \max_{p_{j} \in [\frac{\gamma}{2}, 1 - \frac{\gamma}{2}]} \Big( \frac{(1- \gamma)p_{j} x_{+} + \frac{p_{j} \gamma}{2}}{(1- \gamma)(p_{j} x_{+} + (1 - p_{j})x_{-}) + \frac{\gamma}{2}} \Big) \\
    &= \max_{x_{+}  \in [0, 1]} \max_{p_{j} \in [\frac{\gamma}{2}, 1 - \frac{\gamma}{2}]} \Big( \frac{(1- \gamma)p_{j} x_{+} + \frac{p_{j} \gamma}{2}}{(1- \gamma)p_{j} x_{+} + \frac{\gamma}{2}} \Big) \\
    &=  \max_{p_{j} \in [\frac{\gamma}{2}, 1 - \frac{\gamma}{2}]} \Big( \frac{(1 - \gamma)p_{j} + \frac{p_{j} \gamma}{2}}{(1 - \gamma)p_{j} + \frac{\gamma}{2}} \Big) \\
    &= \frac{(1 - \gamma)(1 - \frac{\gamma}{2}) + \frac{\gamma}{2}(1 - \frac{\gamma}{2})}{(1 - \gamma)(1 - \frac{\gamma}{2}) + \frac{\gamma}{2}} \\
    &= \frac{(1 - \frac{\gamma}{2})^{2}}{(1 - \frac{\gamma}{2})^{2} + (\frac{\gamma}{2})^{2}}
\end{align*}

Thus, $P_{s, lg}\big(\frac{1}{2}, \rhohat_{\pm}\big) \leq p_{j+1}^{*} \leq \frac{(1 - \frac{\gamma}{2})^{2}}{(1 - \frac{\gamma}{2})^{2} + (\frac{\gamma}{2})^{2}}$. Finally, we note that the bound on $p_{j+1}^{*}$ is tight and is achieved when $p_{j} = 1 - \frac{\gamma}{2}$ and $\rhohat_{\pm}^{(j)} = (1 - \gamma) \ketbra{\pm \frac{\pi}{4}} + \frac{\gamma}{2} \mathbb{I}$.

\end{document}